%% file: main.tex
\documentclass[10pt,twocolumn,letterpaper]{article}

\usepackage[pagenumbers]{CVPR2025-v3.1-latex/cvpr} %

\input{preamble}

\definecolor{cvprblue}{rgb}{0.21,0.49,0.74}
\usepackage[pagebackref,breaklinks,colorlinks,allcolors=cvprblue]{hyperref}

\def\paperID{11446} %
\def\confName{CVPR}
\def\confYear{2025}

\title{\ourTitle}

\author{Nam Anh Dinh\\
University of Chicago\\
\and
Itai Lang\\
University of Chicago\\
\and
Hyunwoo Kim\\
University of Chicago\\
\and
Oded Stein\\
University of Southern California\\
\and
Rana Hanocka\\
University of Chicago\\
}
\begin{document}

\ifsuppsubmit
    \input{sections/X_suppl}
    {
        \small
        \bibliographystyle{CVPR2025-v3.1-latex/ieeenat_fullname}
        \bibliography{references.bib}
    }

\else

    \twocolumn[{%
    \renewcommand\twocolumn[1][]{#1}%
    \maketitle
    \input{figures/teaser/teaser.tex}

    }]
    
    \input{sections/0_abstract.tex}

    \input{sections/1_introduction.tex}

    \input{sections/2_related_work.tex}

    \input{sections/3_method.tex}

    \input{sections/4_experiments.tex}

    \input{sections/5_conclusion.tex}

    {
        \small
        \bibliographystyle{CVPR2025-v3.1-latex/ieeenat_fullname}
        \bibliography{references.bib}
    }

    \ifarxivsubmit
        \clearpage
        \input{sections/X_suppl}
    \else
    \fi
\fi

\end{document}

%% file: preamble.tex
\usepackage{multirow}
\usepackage{wrapfig}

\newif\ifarxivsubmit
\arxivsubmittrue

\ifarxivsubmit
\else
    \usepackage[accsupp]{axessibility}
\fi

\newif\ifcamreadydraft
\camreadydraftfalse

\newif\ifdraft
\draftfalse

\newif\ifsuppsubmit
\suppsubmitfalse

\newif\ifsigsubmit
\sigsubmitfalse

\ifsigsubmit
\usepackage{cleveref}
\usepackage{xspace}

\makeatletter
\DeclareRobustCommand\onedot{\futurelet\@let@token\@onedot}
\def\@onedot{\ifx\@let@token.\else.\null\fi\xspace}

\def\eg{\emph{e.g}\onedot} 
\def\ie{\emph{i.e}\onedot} 
 
 \def\vs{\emph{vs}\onedot}

\makeatother

\else
\fi

\ifdraft 
\newcommand{\rana}[1]{{\bfseries \scriptsize \color{purple} Rana: #1}}
\newcommand{\oded}[1]{{\bfseries \scriptsize \color{green} Oded: #1}}
\newcommand{\itai}[1]{{\bfseries \scriptsize \color{blue} Itai: #1}}
\newcommand{\namanhTODO}[1]{{\color{red}\textbf{namanh TODO: }#1}}
\newcommand{\todo}[1]{{\color{red}#1}}

\newcommand{\il}[1]{{\color{blue}#1}}
\newcommand{\brian}[1]{{\color{orange}#1}}

\else
\newcommand{\rana}[1]{}
\newcommand{\oded}[1]{}
\newcommand{\itai}[1]{}
\newcommand{\namanhTODO}[1]{}
\newcommand{\brian}[1]{{\color{black}#1}}
\newcommand{\todo}[1]{}

\newcommand{\il}[1]{{\color{black}#1}}
\fi

\ifcamreadydraft
\newcommand{\namanhCAMRDY}[1]{{\color{green!30!black!70}#1}}

\else
\newcommand{\namanhCAMRDY}[1]{#1}

\fi

\DeclareMathOperator{\SO}{SO}
\DeclareMathOperator*{\argmin}{argmin}

\def\AA{\mathcal{A}}
\def\MM{\mathcal{M}}
\def\VV{\mathcal{V}}
\def\FF{\mathcal{F}}
\def\UU{\mathcal{U}}
\def\RR{\mathcal{R}}
\def\NN{\mathcal{N}}

\def\uu{\mathsf u}
\def\vv{\mathsf v}
\def\xx{\mathbf x} %
\def\ee{\mathsf e}

\def\Mm{\mathsf M}
\def\Rr{\mathsf R}

\def\ourmethod{Geometry in Style}
\def\ourTitle{\ourmethod: 3D Stylization via Surface Normal Deformation}
\def\ourtechnique{dARAP}

%% file: sections/X_suppl.tex
\clearpage
\appendix
\setcounter{page}{1}
\maketitlesupplementary
\setcounter{figure}{11}
\setcounter{table}{1}
\setcounter{equation}{7}

These sections provide additional results and more information on our deformation method. \cref{sec:supp:qualresults} presents qualitative results and comparisons in addition to those from the paper. In \cref{sec:supp:clipsim}, we show an evaluation of CLIP similarity to the prompt. In \cref{sec:supp:ablation}, we perform an ablation on the rotation-finding method and show the regularization significance of the Procrustes local step. In \cref{sec:supp:csd}, we explain in detail the Cascaded Score Distillation (CSD) semantic loss, introduced by \citet{decatur2024paintbrushcsd}, and provide the hyperparameters and configuration we used. In \cref{sec:supp:additional-quant-details} we provide extra clarifications on the quantitative evaluation from the main paper. Finally, in \cref{sec:supp:runningtime}, we show a running time comparison of the global step solves of dARAP and NJF \cite{aigerman2022neural}.

\section{Additional Qualitative Results}
\label{sec:supp:qualresults}

\begin{figure}[b]
    \centering
    \includegraphics[width=0.99\linewidth]{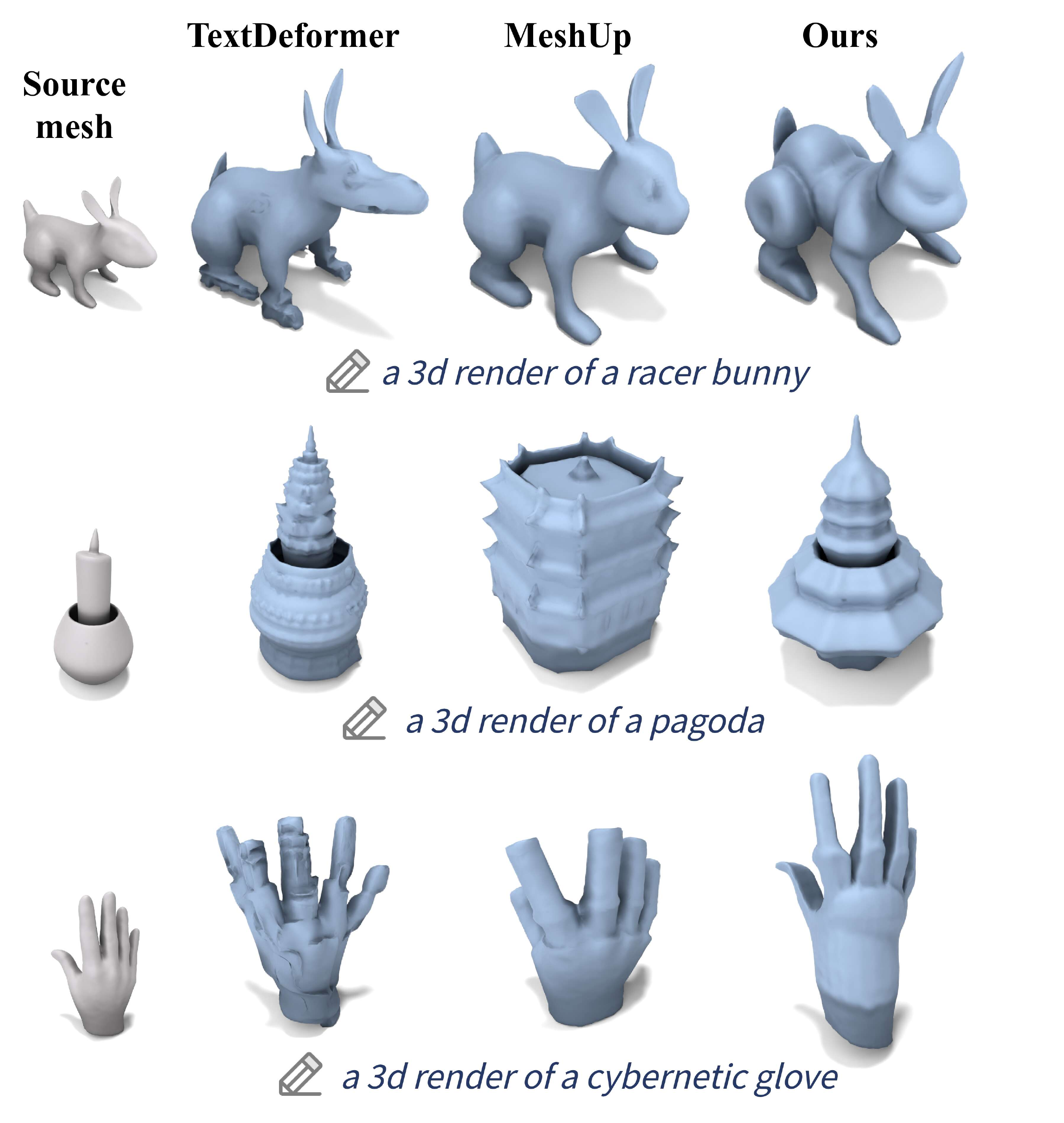}
    \caption{\textbf{More qualitative comparisons with MeshUp and TextDeformer.} Note that TextDeformer contains artifacts and noisy surface detail, and MeshUp either fails to attain the prompt style, or deforms the shape beyond recognition and loses the source shape's identity. Our method attains the target style while preserving the source shape's identity.
}
    \label{fig:supp-more-qualcomparison}
\end{figure}

In \cref{fig:supp-more-qualcomparison} we present additional qualitative comparisons of our method against MeshUp and TextDeformer. 
In the \emph{racer bunny} prompt, our method achieves the target style with most fidelity. In the \emph{pagoda} prompt, TextDeformer produces artifacts and achieves less satisfactory style; MeshUp achieves the target style but aggressively transforms the shape beyond recognition from the source, losing its identity. Our method achieves the \emph{pagoda} style and retains the original vessel-and-column configuration. In the \emph{cybernetic glove} example, our method achieves the target style cleanly and preserves the slender proportions of the source hand.

\begin{figure}
    \centering
    \includegraphics[width=0.99\linewidth]{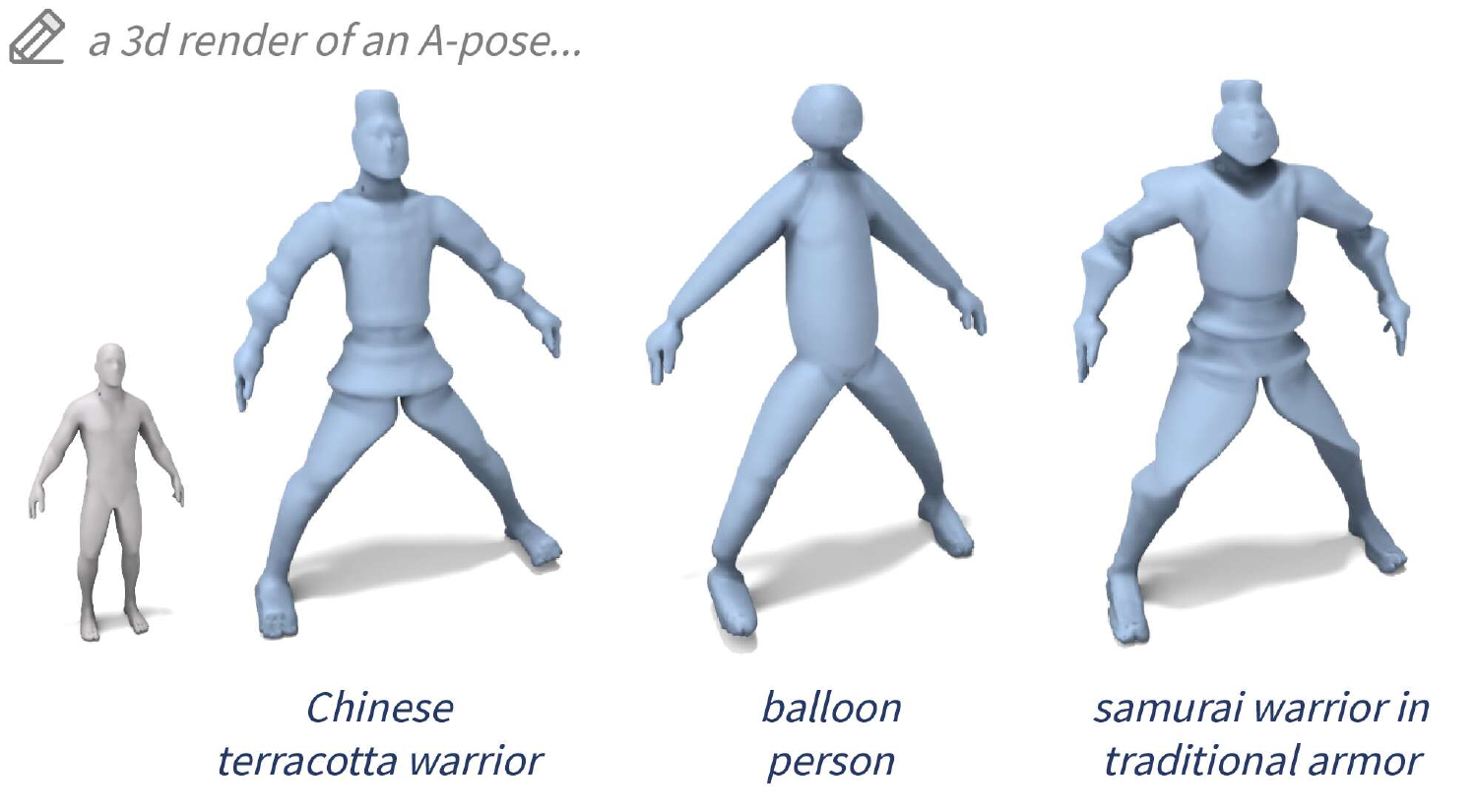}
    \caption{Another example of deforming the same mesh towards different text-specified styles.}
    \label{fig:same-shape-aposeonly}
\end{figure}

In \cref{fig:same-shape-aposeonly}, we add to Fig. 5 in the main paper with another example showing the same shape deformed with different style prompts, here a person shape. All three examples show salient features of the style but preserve the slender proportions of the original shape.

In \cref{fig:gallery-supp} we present additional results of our method on select quadruped animal meshes from the SMAL model \cite{Zuffi2017SMAL} (the image-fitting result meshes presented in their work) paired with diverse prompts.

\section{CLIP Similarity to Prompt}
\label{sec:supp:clipsim}

We computed the CLIP similarity between rendered images of the deformed mesh and the stylization text prompt. Our evaluation on the 20-shape set (the same set evaluated quantitatively in the main paper) using 16 views per mesh shows (\cref{tab:clipsim}) that we obtain better semantic similarity to the specified style prompt compared to MeshUp \cite{kim2024meshup} and TextDeformer \cite{gao2023textdeformer}.

\begin{table}
    {\small
\begin{tabular}{rccc}
\toprule
& TextDeformer & MeshUp & \textbf{Ours} \\
\midrule
ViT-B/16 CLIP sim. ($\uparrow$) & 0.650  & 0.653 & \textbf{0.655} \\
\bottomrule
\end{tabular}
}
    \caption{{Our method achieves better CLIP similarity to the prompt than TextDeformer and MeshUp.}}
    \label{tab:clipsim}
\end{table}

\section{Ablations}
\label{sec:supp:ablation}

\paragraph{Rotation-finding method. }

\begin{figure}
    \includegraphics[width=0.99\linewidth]{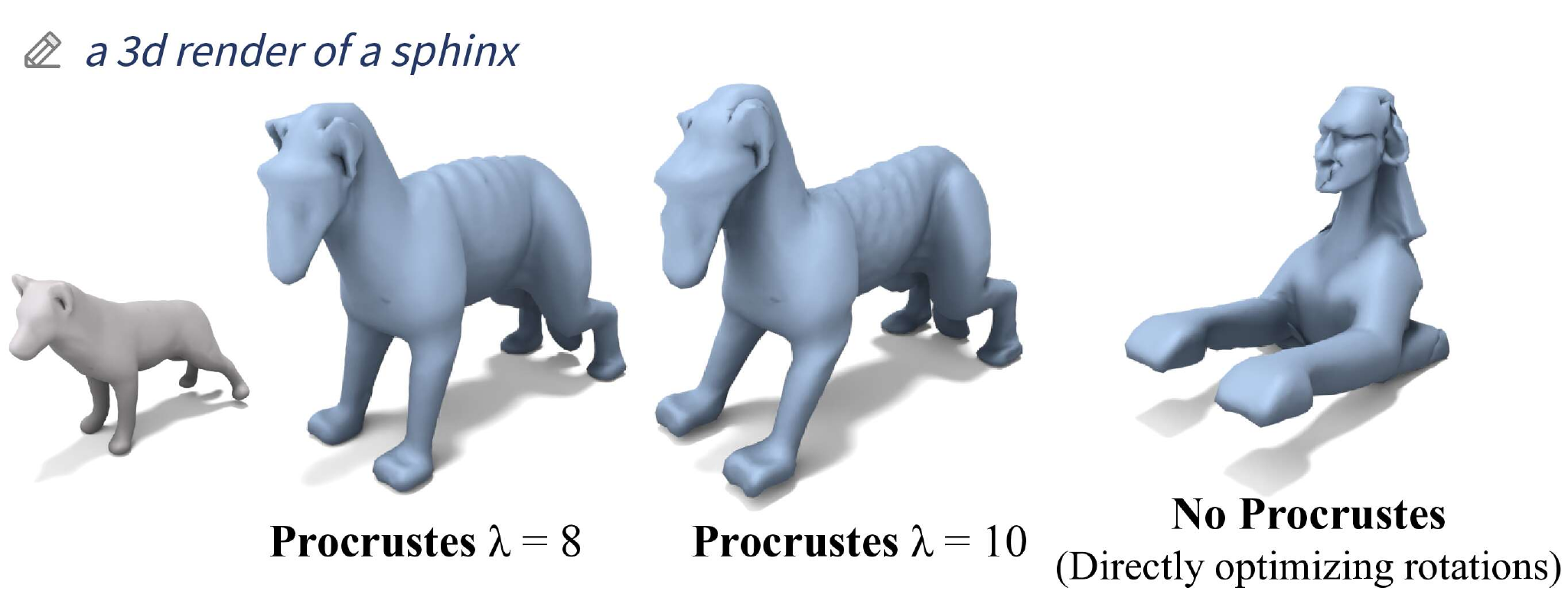}
    \captionof{figure}{\textbf{Rotation ablation.} Our Procrustes solve preserves identity even at different values of $\lambda$ (tunable to taste). Directly predicting rotations is not restrictive enough and severely changes shape identity.}
    \label{fig:ablation-localstep}
\end{figure}

\begin{figure*}[]
    \centering
    
    \ifarxivsubmit
        \includegraphics[width=0.9\linewidth]{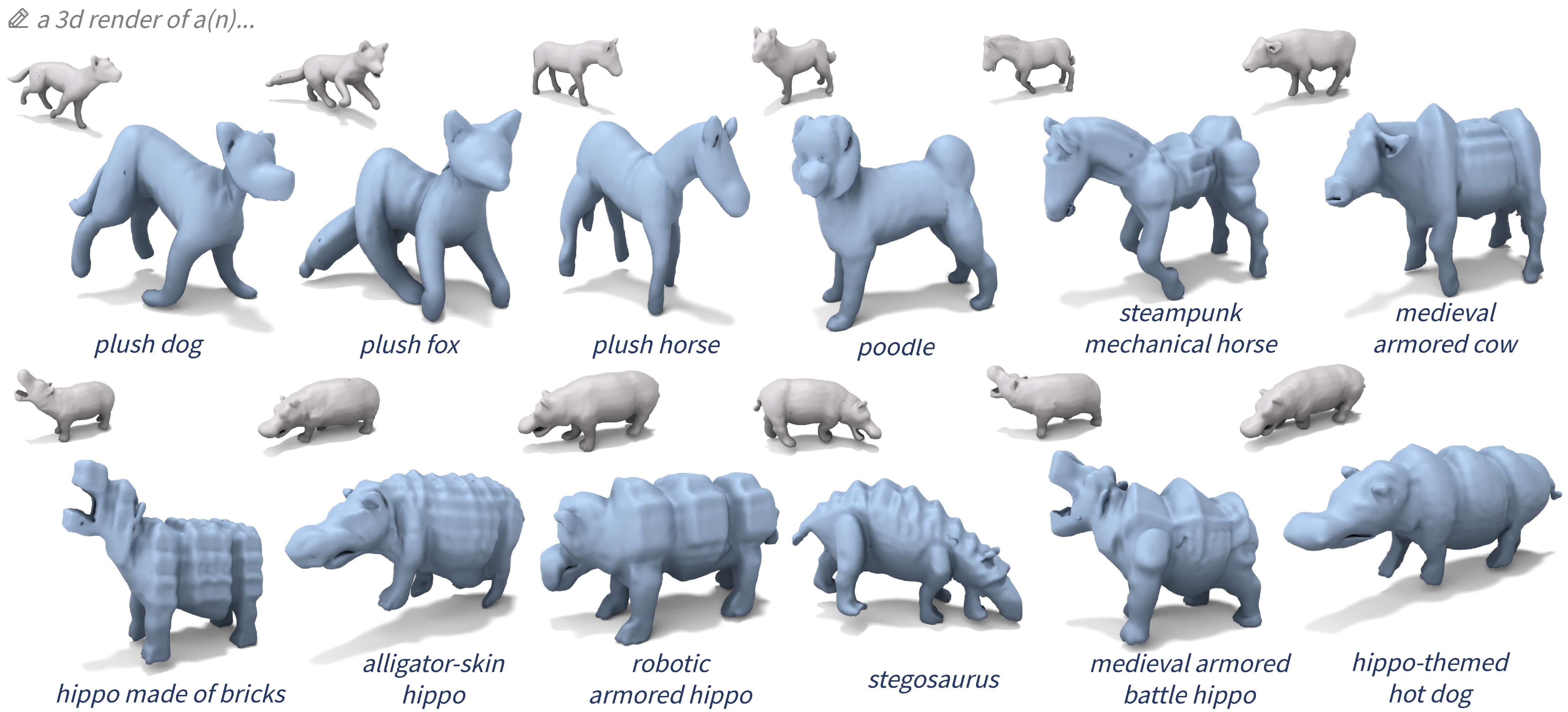}
    \else
        \includegraphics[width=0.99\linewidth]{figures/supp/cvpr25_figures-drawio-gallery-supp.pdf}
    \fi
    \caption{\textbf{Diverse prompts on SMAL \cite{Zuffi2017SMAL} quadruped animals.} For a variety of animals, the deformation imparts the requested style but preserves the general pose. Even in cases where the prompt's identity is different (\eg \emph{poodle, stegosaurus, hot dog}), our method works with the existing geometry and keeps its identity, effectively stylizing the original shape to look like the target object rather than overriding the source shape and growing a distinct body.}
    \label{fig:gallery-supp}
\end{figure*}

In \cref{fig:ablation-localstep}, we compare choices of the local rotation-finding method. Our local step using a Procrustes solve, which finds a rotation matrix for each vertex given a target normal, is inherently regularized by virtue of finding a best-fit rotation for not only the source normal but also a neighborhood of edge vectors (Fig. 4).

Without this Procrustes solve, one might \emph{directly optimize} per-vertex rotations (\eg using the continuous $3\times 2$ matrix representation described in \citet{zhou2019continuityrot3x2}). However, we show in \cref{fig:ablation-localstep} that this, without identity regularization loss, is far less restrictive than our target-normal-based local step and loses the identity of the source mesh, undesirable for our goal of stylization.

Further, our Procrustes solve hyperparameter $\lambda$ can be chosen (for optimization) to add additional detail yet still preserve identity to taste. \cref{fig:ablation-localstep} shows that the stylization is visible yet preserves the identity of the source dog mesh for both $\lambda=8$  and $\lambda=10$. In particular, the squared feet and sharper angle of the neck are prominent in both, with $\lambda=8$ preserving the front leg pose better, while $\lambda=10$ having stronger sphinx-like feet. We use $\lambda=8$ in the main results as it strikes a good balance of identity preservation and strength for most meshes, but this can be tuned on a per-mesh basis. Additionally, as shown in Fig. 8, $\lambda$ can also be changed at inference/application time to further adjust a pre-optimized deformation's strength.

\section{Cascaded Score Distillation (CSD)}
\label{sec:supp:csd}

\subsection{Method}
\begin{figure*}[t]
    \centering
    \ifarxivsubmit
        \includegraphics[width=0.82\linewidth]{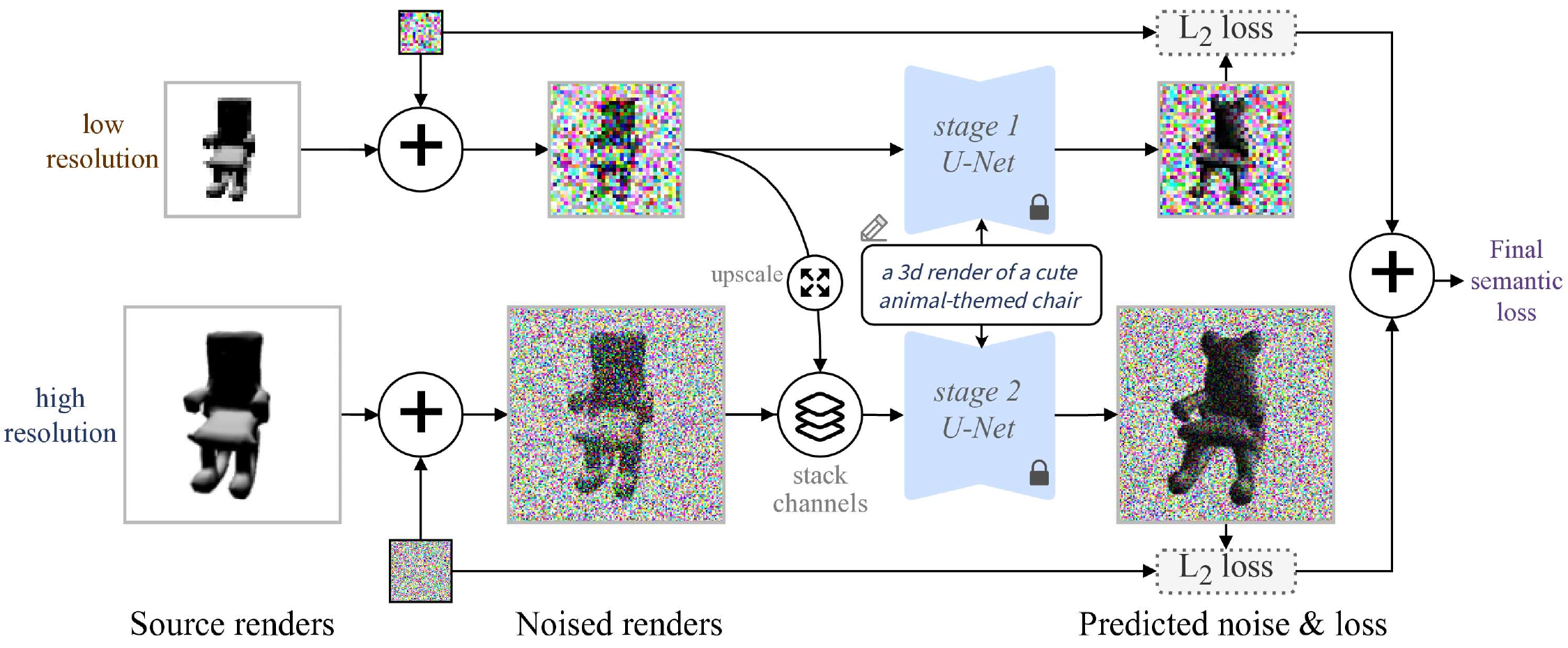}
    \else
        \includegraphics[width=0.9\linewidth]{figures/supp/cvpr25_figures-drawio-suppcsd.pdf}
    \fi
    \caption{\textbf{Overview illustration of the Cascaded Score Distillation semantic loss.} Two-stage CSD uses two pretrained denoising U-Nets to predict the noise for the noised renders at their corresponding resolutions. The higher resolution stage incorporates an upscaled noised render from the low-resolution branch.}
    \label{fig:suppcsd}
\end{figure*}

In this section, we briefly describe how Cascaded Score Distillation is used to optimize the target vertex normals $\hat \UU = \{\hat \uu_k \mid k \in \{1\dots|\VV|\}\}$. We first provide an overview of how Score Distillation Sampling (SDS) can be applied to optimize a deformation quantity (\eg jacobians, or target normals in our case) using diffusion models, and extend the concept to show how we use Cascaded Score Distillation as our guidance. See \cref{fig:suppcsd} for an illustrative overview.

 To stochastically optimize parameters (target normals in our case) with respect to a pre-trained 2D diffusion model, \citet{poole2022dreamfusiontextto3dusing2d} proposed Score Distillation Sampling (SDS), where given a rendered image $\mathbf{z}$ and a text condition $y$, the objective is to minimize the L2 loss 
\begin{equation}
\mathcal{L}_{\text{Diff}}(\omega, \mathbf{z}, y, \epsilon, t)=w(t)\left\|\epsilon_\omega\left(\mathbf{z}_t, y, t\right)-\epsilon\right\|_2^2,
\end{equation}
which is the L2 difference between the sampled noise $\epsilon~\sim~\mathcal{N}(0, \mathbf{I})$ added to the image, and the noise $\epsilon_\omega$ predicted by a denoising U-Net $\omega$ at some timestep $t$, sampled from a uniform distribution $t \sim U(0,1)$.
Here, $w(t)$ is a weighting term, and \(\mathbf{z}_t\) is the rendered image, noised according to the sampled timestep $t$. 
To compute the gradient of the optimizable parameters, which in our case is the set of all target normals $\uu_k$ with respect to the loss $L_\mathrm{Diff}$, it has been shown that the gradients of the U-Net can be omitted for efficient computation \cite{poole2022dreamfusiontextto3dusing2d, scorejacobian}, giving %
\begin{equation}
\nabla_{\uu_k} \mathcal{L}_{\mathrm{SDS}}(\phi, \mathbf{z}, y, \epsilon, t) = w(t) \left( \epsilon_{\omega}(\mathbf{z}_t, y, t) - \epsilon \right) \frac{\partial \mathbf{z}_t({\uu_k})}{\partial \uu_k}.
\label{eq:sds}
\end{equation}

$\frac{\partial \mathbf{z}_t({\uu_k})}{\partial \uu_k}$ can be obtained by backpropagating the gradient from the rendered images through our fully differentiable pipeline, and using $\nabla_{\uu_k}$ we can optimize the target normals with text-to-image diffusion models.

Cascaded Score Distillation is an extension of SDS that allows our parameters to be optimized with additional guidance from super-resolution models. While SDS only approximates the gradients from the first stage, base diffusion model, CSD utilizes the super-resolution models (usually 2nd or 3rd stage models), which are additional models used to upsample and fine-tune low-resolution images generated by the base model. This additional guidance from high-resolution modules allows for a more fine-grained generation of stylistic details in our results, making it an appropriate choice for our 

To use Cascaded Score Distillation as our guidance, we add the gradient,

\begin{equation}
\begin{split}
\nabla_{\uu_k} \mathcal{L}_{\mathrm{CSD}_i}&(\phi^i, z^i, z^{i-1}, y) =\\
&w(t) \left( \epsilon_{\phi^i}(z_t^i, t, z_s^{i-1}, s, \mathbf{y}) - \epsilon^i \right)\frac{\partial \mathbf{z}_t({\uu_k})}{\partial \uu_k},
\end{split}
\end{equation}
where the idea is to use an image upsampled at the resolution of the $i^{th}$ stage module, $z^i$, along with the image of the resolution of the previous $i-1^{th}$ stage module, $z^{i-1}$, and noise them respectively at timestep $t$ and $s$ to use them as inputs to the $i^{th}$ stage high-resolution module. Likewise for SDS, we omit the expensive computation of the gradient through the U-Net and calculate the gradient with respect to our target normals by backpropagating through our differentiable pipeline. 

Combining the gradient from SDS with CSD gives us the following final gradient, 
\begin{equation}
\label{eq:csd}
\begin{split}
\nabla_{\uu_k} \mathcal{L}_\mathrm{CSD}&(\phi, \mathbf{z}, y, \epsilon, t)
=\\ &\alpha^1 \nabla_{\uu_k} \mathcal{L}_\mathrm{SDS}
+ \sum_{i=2}^N \alpha^i 
    \nabla_{\uu_k} \mathcal{L}_\mathrm{CSD_i}(\phi^i, z^i, z^{i-1}, \mathbf{y})
\end{split}
\end{equation}
where $\alpha^i$ are the user-defined weights for each gradient.

\subsection{Configuration Details}
\paragraph{CSD settings. }
Recall that an epoch in our optimization pipeline consists of a batch of randomly sampled views (we use a view batch size of 8) fed to CSD. Apart from the base model for SDS, we only use one upscaling stage of DeepFloyd IF (\ie $N=2$ in  \cref{eq:csd}) The weight $\alpha^2$ of the CSD second-stage is linearly ramped up from 0 to 0.2 over the course of 1000 epochs, then 0.2 to 0.3 over 750 epochs, remaining at 0.3 for the rest of the epochs. The weight $\alpha^1$ for SDS is a constant $1.0$. We use a classifier-free guidance weight of 100 as in MeshUp \cite{kim2024meshup}.

For each batch of rendered views we compute the CSD loss, backpropagate, and perform a gradient descent update on $\hat\UU$ \emph{twice} before recomputing the deformed shape and re-rendering it for the next epoch/batch of views. This technique was also used in the official MeshUp implementation and empirically leads to sharper deformations.
\paragraph{View sampling settings. }
The renders sample from a range of views around the mesh: a full azimuth range of $0^\circ$ to $360^\circ$, an elevation range of $0^\circ$ to $60^\circ$ (or $30^\circ$ for tall and slim shapes such as humans), a distance range of $2.5$ to $3.0$ for most shapes (or $1.4$ to $2.6$ for tall and slim shapes such as humans), and a fixed FOV of $60^\circ$.

\section{Additional Quantitative Evaluation Details}
\label{sec:supp:additional-quant-details}

We provide additional details and explanations about the quantitative evaluation reported in Tab. 1 in the main paper. For an informative comparison, we normalize the source shape to fit a side-2 cube centered at origin, and rescale the deformed shape to share the same axis-aligned bounding box diagonal length as that of the normalized source shape. We use this pre- and post-normalization as it is the same scheme used before and after deformation in MeshUp and our method. This normalization means the face area ratio will measure \emph{two effects}: the first is any localized face area distortion to accommodate a deformation, and the second is if the deformation (before normalizing) significantly enlarges the shape's bounding box even if there is no significant localized distortion  (\eg when a limb is rotated to spread out further). When the latter case happens, after the normalization, the area ratios will be smaller than 1 due to a global rescaling down to fit the original bounding box extent, and vice versa. 

\begin{figure}
    \centering
    \ifarxivsubmit
        \includegraphics[width=0.8\linewidth]{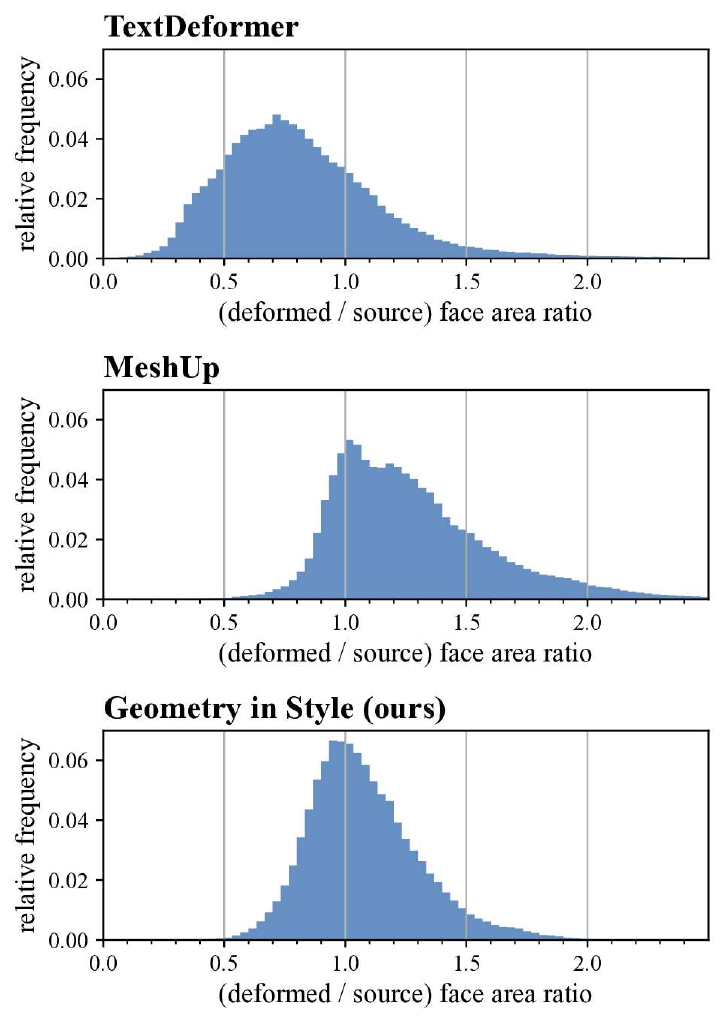}
    \else 
        \includegraphics[width=0.99\linewidth]{figures/supp/prerenderedfig-face-area-ratio-histo.pdf}
    \fi
    \caption{\textbf{Distribution of (deformed / source) face area ratios.} $n=306300$ faces across 20 source-deformed mesh pairs.}
    \label{fig:face-area-ratio-histo}
\end{figure}

In addition to the average and standard deviation of the area ratio reported in the main body, we further show the distribution of the area ratio in \cref{fig:face-area-ratio-histo}. Interestingly, we see that the distribution for TextDeformer \cite{gao2023textdeformer} is mainly centered around values smaller than 1, suggesting that TextDeformer tends to enlarge the source shape's extents as a whole (resulting in a global shrinkage upon normalizing), while for MeshUp \cite{kim2024meshup} the behavior is the other way around. In contrast, the distribution for our method is cleanly located around a ratio of value 1, indicating better triangle area and bounding box preservation compared to the other methods.

\section{Running Time}
\label{sec:supp:runningtime}

We compare the running time of \ourtechnique{} against the equivalent in NJF \cite{aigerman2022neural}. Since faces approximately outnumber vertices 2:1 on simplicial surfaces, our vertex-based global solve is faster than NJF's face jacobian-based solve (0.0507s \vs 0.0755s average). Even combined with our local step, the two methods are comparable in run time; see the supplementary material for measurement details.

We compare in \cref{tab:runtime} the running time of \ourtechnique{} against the equivalent in NJF \cite{aigerman2022neural}. Averaged over 10 runs on a mesh with 20708 faces and 10356 vertices on a GTX 1660Ti GPU, excluding the precomputations of both methods, our Poisson system construction and solution is faster than that of NJF, owing to the fact that faces usually outnumber vertices 2:1 on simplicial surfaces. While NJF does not involve a local step, even coupled with the local step, our local-global \ourtechnique{} altogether has a comparable running time to an NJF Poisson solve.

\begin{table}
    \centering
    \begin{tabular}{lcc|c}
    \toprule
    Method & Local step & Global step & Total\\
    \midrule
    NJF~\cite{aigerman2022neural} Poisson  & --- & 0.0755s & 0.0755s \\
    \ourtechnique{} (\textbf{ours}) & 0.0412s & 0.0507s & 0.0918s \\
    \bottomrule
    \end{tabular}
    \caption{\textbf{Running time comparison against NJF.} Since faces outnumber vertices 2:1 on simplicial surfaces, our vertex-based global solve is faster than NJF's face jacobian-based solve. Combined with our local step, the two methods are on par in run time.}
    \label{tab:runtime}
\end{table}

%% file: figures/teaser/teaser.tex
\begin{center}
    \centering
    \vspace{-22pt}
    \includegraphics[width=0.99\linewidth]{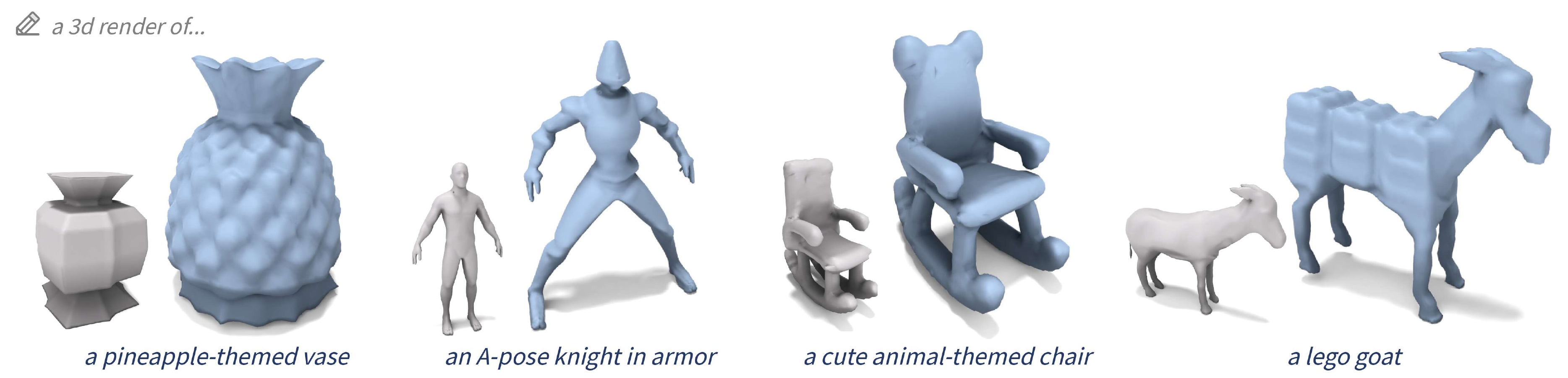}
    \vspace{-5pt}
    \captionof{figure}{Our method deforms a source shape (gray) into a text-specified \textit{semantic style} (blue). While the deformations are expressive, they \textit{preserve the identity} of the original shape.}
    \label{fig:teaser}
\end{center}

%% file: sections/0_abstract.tex
\ifsigsubmit
\else
\begin{abstract}
\fi

We present \textit{Geometry in Style}, a new method for identity-preserving mesh stylization. Existing techniques either adhere to the original shape through overly restrictive deformations such as bump maps or significantly modify the input shape using expressive deformations that may introduce artifacts or alter the identity of the source shape. In contrast, we represent a deformation of a triangle mesh as a target normal vector for each vertex neighborhood.
The deformations we recover from target normals are expressive enough to enable detailed stylizations yet restrictive enough to preserve the shape's identity. We achieve such deformations using our novel differentiable As-Rigid-As-Possible (\ourtechnique{}) layer, a neural-network-ready adaptation of the classical ARAP algorithm which we use to solve for per-vertex rotations and deformed vertices.
As a differentiable layer, \ourtechnique{} is paired with a visual loss from a text-to-image model to drive deformations toward style prompts, altogether giving us \ourmethod{}. Our project page is at {\footnotesize\url{https://threedle.github.io/geometry-in-style}}.

\ifsigsubmit
\else
\end{abstract}
\fi

%% file: sections/1_introduction.tex
\section{Introduction} \label{sec:introduction}

Semantically deforming triangle meshes is a basic task in 3D surface modeling.
A common paradigm for shape creation is to take a base 3D object and deform it to sculpt a desired shape. For example, a human artist creates an intricate surface by starting with a simple generic version of an object (from a shape library, or quickly sketched),  and successively deforms parts of the object like a sculptor modeling clay.

Recently, machine learning pipelines have adopted deformations as a strategy for neural shape manipulation~\cite{jakab2021keypointdeformer,aigerman2022neural,shechter2022neuralmls}. 
However, existing learning-based methods do not always fully mimic the classical approach: they either make small, surface-level changes for simple stylization~\cite{michel2022text2mesh,hertz2020deep}, or they generate large deformations that destroy the identity of the base object~\cite{gao2023textdeformer,kim2024meshup}.
Our method, \emph{\ourmethod{}}, uses text prompts to generate large deformations that create unique shapes while preserving the \emph{identity} of the base shape (see \cref{fig:teaser}) -- much like a human sculptor would.

\input{figures/gallery/gallery_pdf.tex}

We posit that existing learning methods' failure to generate the kinds of large, identity-preserving deformations that a human artist would is due to not employing the right representation for the space of deformations.
Existing methods either (a) adhere to the original shape through overly restrictive deformations, such as bump maps~\cite{michel2022text2mesh, hertz2020deep};
or they (b) significantly modify the input shape using expressive deformations \il{based on gradient fields} that may introduce artifacts or destroy the identity of the original shape~\cite{gao2023textdeformer, kim2024meshup}.

In this work, a deformation of a triangle mesh is represented by target normals for vertex neighborhoods. We recover a deformation from this representation using our differentiable As-Rigid-As-Possible method (\ourtechnique{}), whose formulation optimizing for local rigidity yields detailed and salient deformations that are nonetheless restrictive enough to preserve the identity of the base shape. 
\ourtechnique{} locally rotates each vertex neighborhood individually to fit its normal to a desired target normal, and follows this local rotation with a global step that finds a global deformation that best fits all individual rotated neighborhoods.
Crucially, \ourtechnique{} is differentiable, and can be used as a layer in a neural network.
This is achieved by replacing classical ARAP's iteration of local and global steps until convergence (impractical to backpropagate through) with \ourtechnique{}'s easily differentiable use of a single local and global step.
Where classical ARAP needs many iterations to converge in deformation tasks with fixed target vertex positions, \ourtechnique{}'s use with target normals inside the \emph{\ourmethod{}} method (where \ourtechnique{} is run once per iteration of a larger gradient descent optimization problem) achieves high-quality deformations from only a single iteration. 

We use \ourtechnique{} together with a visual loss from a text-to-image model (T2I) which drives our deformation
to arrive at \emph{\ourmethod{}}.  Our visual loss leverages a cascaded T2I model to achieve high-fidelity deformations~\cite{decatur2024paintbrushcsd}, allowing the use of a user-specified text prompt to deform any base shape  into a stylized object without any dedicated 3D supervision data.
As such, our method allows the application of a wide variety of styles, indicated intuitively by text prompts, to a wide variety of shapes (\cref{fig:teaser,fig:gallery}). Our stylization can manifest as different types of geometric manipulation, such as local surface texture, as well as global low-frequency deformations. Additionally, we show that our method offers control over the deformation result, where the user can easily change the strength of the stylization effect even after optimization.
We contrast our method with recent deformation techniques and find that it can better achieve the target style with a lower surface distortion.

\smallskip
In this work, we present:
\begin{itemize}
    \item \emph{\ourtechnique{}}, a differentiable neural network layer that deforms a triangle mesh to specified target normals; and
    \item \emph{\ourmethod{}}, an identity-preserving shape deformation method from user text input using target normals as a representation of the space of deformations;
\end{itemize}
We achieve high-quality stylization of shapes through deformation that is faithful to the identity of the input shape with a simple and easy-to-implement framework.

%% file: figures/gallery/gallery_pdf.tex
\begin{figure*}
    \centering
    \includegraphics[width=0.90\linewidth, trim=0 0 0 0, clip]{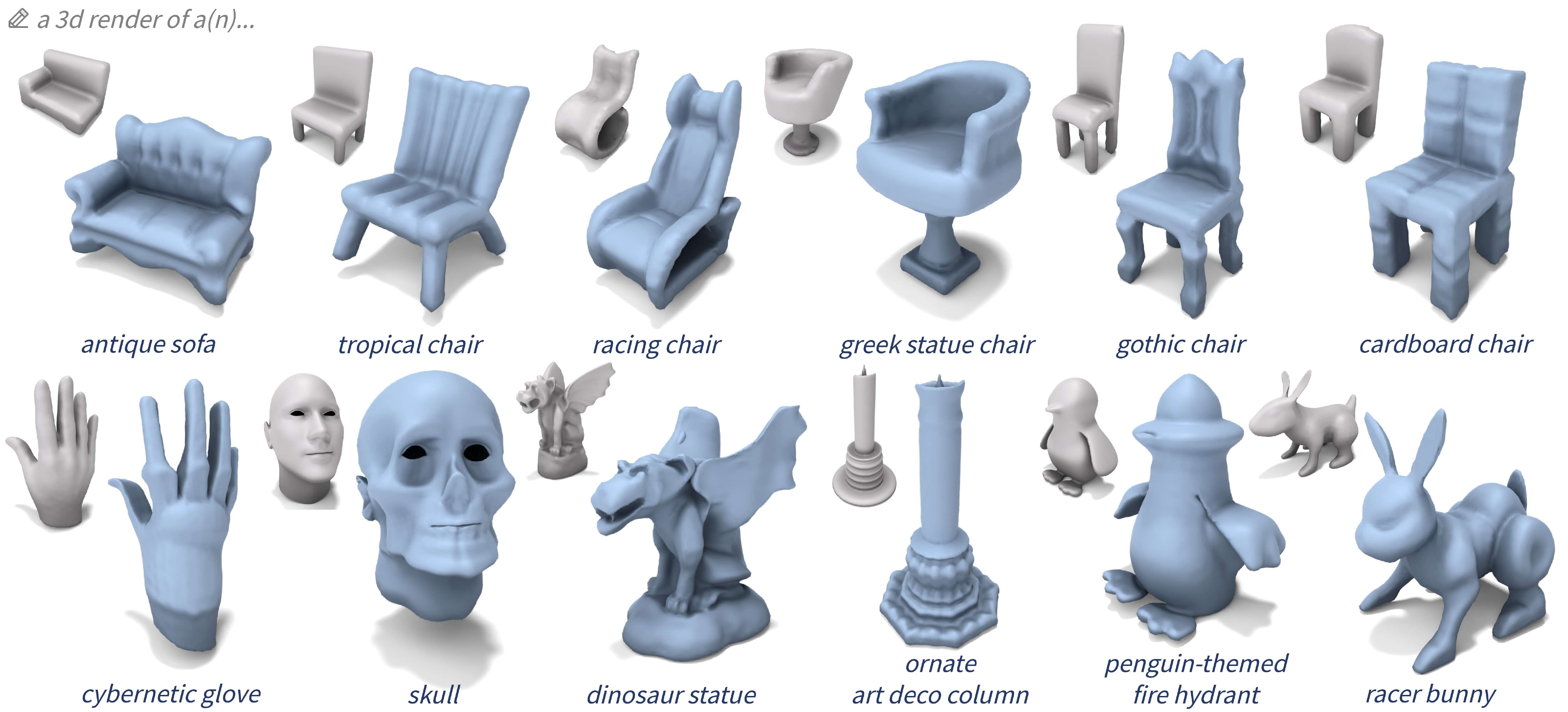}
    \vspace{-5pt}
    \caption{\textbf{Style diversity.} Our method is capable of deforming various input meshes towards a variety of text-specified styles. The style can be manifested as fine geometric details, like in the \emph{ornate art deco column}, or as low-frequency deformations, such as the joints of the \emph{cybernetic glove}. Our method retains the structural features of the input shape, such as a flat arm on the \emph{antique sofa}. Moreover, the resultant stylizations are in accordance with prompt semantics and part-aware semantics: the folds in the tropical chair are on the seat and backrest as opposed to the legs, the head of the penguin becomes like the top of a fire hydrant, and the racer bunny's thigh turns into the shape of a wheel.}
    \label{fig:gallery}
\end{figure*}

%% file: sections/2_related_work.tex
\section{Related work} \label{sec:related_work}

\begin{figure*}
    \centering
    \includegraphics[width=0.91\linewidth, trim=19 0 0 0, clip]{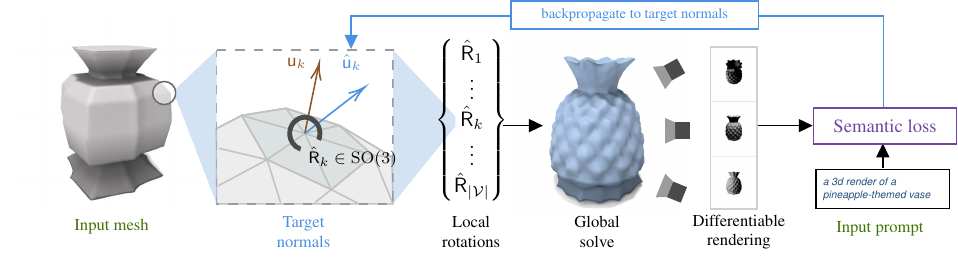}
    \vspace{-5pt}
    \caption{Overview of our stylization pipeline. \ourmethod{} optimizes vertex normals to deform the mesh surface, subject to a stylization text prompt. Using the normals undergoing optimization as a target for our differentiable As-Rigid-As-Possible method (\ourtechnique{}), the \ourtechnique{} local step computes a rotation matrix per vertex; we then obtain the deformed surface via our \ourtechnique{} global solve. Then, we utilize a differentiable renderer and a diffusion model-based semantic loss to guide the normals being optimized towards a deformation matching the desired style prompt.}
    \label{fig:overview}
\end{figure*}

\subsection{Classical Deformation Representations}
Our deformation technique builds on the extensive literature on variational surface deformation and mapping methods \cite{botschsorkine08deformsurvey}. Methods in this family commonly compute a local differential deformation quantity, such as a per-element rotation, normal, jacobian matrix, or differential coordinates, subject to modeling constraints such as handles or cages. The vertex positions are then recovered using a global linear solve, usually derived from the least-squares minimization of energy that encodes the desired local properties. %

Important classical deformation approaches include Laplacian surface editing \cite{sorkine2004laplacian}, gradient field mesh editing \cite{yu2004poisson}, or skinning-based approaches \cite{Fulton:LSD:2018, skinningcourse:2014}.
A seminal work in this field is ``as-rigid-as-possible'' (ARAP) shape modeling which regularizes the local deformations of a surface to be rigid \cite{sorkine2007arap}. This approach promotes local rigidity with smooth and detail-preserving solutions, with various downstream applications including editing, parameterization, registration, shape optimization, and simulation \cite{igarashi2005rigid,chao2010spokesandrims,liu2008arap,zollhofer2014realnonrigid,bouaziz2012shapeup,liu2013fast}. 
\namanhCAMRDY{We adopt ARAP as the basis for our geometric stylization pipeline's deformation method. This is similar to recent neural methods that have also taken advantage of ARAP in other shape representations \cite{Huang2021arapreg, baieri2024implicitarap, bozic2020neuralnonrigid}, especially the work of \citet{Yan2022_NEO3DF} who incorporate a differentiable ARAP loop into an image-to-3D face reconstruction pipeline. \citet{Yan2022_NEO3DF}'s use of ARAP serves to smoothen an assembly of 3D patches and requires multiple ARAP steps; on the other hand, our dARAP method is meant for optimizations of a deformation quantity (in our case, per-vertex normals) to achieve a desired deformation in a \emph{single} local step-global step pair.}

Often related to such differential deformation methods, the manipulation of surface normals is a cornerstone useful for a variety of applications: shape abstraction \cite{alexa2021polycover}, texture mapping \cite{tarini2004polycube, huang2014l1, zhao2018robust}, mesh parameterization \cite{zhao2020mesh}, generative shape refinement \brian{\cite{li2024craftsmanhighfidelitymeshgeneration}}, and more.
Operating on surface normals has also been particularly core to cubic stylization \cite{zhao2020mesh, fumero2020nonlinear, liu2019cubic} as well as geometric filters \cite{prada2015shockfilter,li2020normalfilternet,zhang2015guidedfilter,liu2018propagatedfilter,zheng2010bilateralfilter}.
Some approaches use an ARAP-like optimization to achieve desired target normals (similar to our goal) for manufacturing \cite{herholz2015approxheightfield,stein2019molds}.

\smallskip
\noindent \textbf{Normal-Driven Shape stylization.}
\citet{liu2021normal} propose a normal-based stylization approach by shape analogies. Given a source object and a sphere-based normal template, modeled as a normal-to-normal function $S^2 \to S^2$, the source shape's normals are locally rotated to match target normals dictated by the template; the deformation is obtained via ARAP solve using these local rotations. Similar to this work, we also use target normal vectors as the driving tool for our deformation. However, we use a text prompt to describe the desired style rather than a geometric exemplar, enabling semantic styles (\eg ``antique'') that are not easily represented by a spherical normal template. Not being tied to a normal template, our deformations are part-aware, \ie different parts with the same source normal do not have to receive the same target normal, and can be stylized differently as can be seen in \cref{fig:teaser,fig:gallery}).

\subsection{Neural Shape Manipulation}
\brian{Following the success of generative methods that optimize 2D representations via text-to-image guidance from diffusion models \cite{DDIM, DDPM, latentdiffusion, karras2022elucidatingdesignspacediffusionbased, song2021maximumlikelihoodtrainingscorebased, song2021scorebasedgenerativemodelingstochastic} or CLIP-based scores \cite{CLIP, zeroshot}, there has been a large body of work using score distillation-based approaches \cite{poole2022dreamfusiontextto3dusing2d, scorejacobian} to achieve 3D generation using 2D diffusion priors. These methods use a variety of shape representations, mostly implicits: Signed Distance Fields (SDF) \cite{park2019deepsdflearningcontinuoussigned, Oleynikova2016SignedDF, chibane2020neuralunsigneddistancefields}, other implicit neural fields  \cite{mescheder2019occupancynetworkslearning3d,chen2019learningimplicitfieldsgenerative, sitzmann2020implicitneuralrepresentationsperiodic, chabra2020deeplocalshapeslearning, gropp2020implicitgeometricregularizationlearning, atzmon2020salsignagnosticlearning} or variations on Neural Radiance Fields (NeRF) and Gaussian splatting \cite{gaussiansplatting,mildenhall2020nerfrepresentingscenesneural,wang2023prolificdreamer,fantasia3d,lukoianov2024scoredistillationreparametrizedddim,liang2023luciddreamerhighfidelitytextto3dgeneration,zero123,zeroshot,clipnerf,zhang2023text2nerf,sweetdreamer,efficientdreamer,seo2023let,magic3d,shi2024mvdreammultiviewdiffusion3d,yi2024gaussiandreamerfastgenerationtext,ma2023geodreamdisentangling2dgeometric,zhao2024animate124animatingimage4d,bahmani20244dfytextto4dgenerationusing, zhu2024hifahighfidelitytextto3dgeneration,qian2023magic123imagehighquality3d, chen2024textto3dusinggaussiansplatting}. Some methods leverage the rapidly growing size of 3D datasets to train text-to-3D models that directly generate 3D representations \cite{jun2023shapegeneratingconditional3d, zhang20233dshape2vecset3dshaperepresentation, nichol2022pointegenerating3dpoint}.}

More relevant to our present work, recent methods have used the strong representation power of neural networks to drive not just generation but also the editing and manipulation of shapes \cite{hertz2020deep, michel2022text2mesh, aigerman2022neural}.  \citet{hertz2020deep}'s network predicts local vertex displacements to match local geometric characteristics of an exemplar shape. The follow-up Text2Mesh \cite{michel2022text2mesh} replaced reference shapes by style text prompts. Similarly, our shape editing is flexibly guided by text, but rather than \textit{raw vertex displacements}, we find rotations of the surface normal and recover an identity-preserving deformation via the dARAP solver.

Other neural deformation and manipulation methods based include data-driven cage deformations \cite{yifan2020neuralcages}, geometric fields for skinning \cite{dodik2024robust}, vector displacement maps \cite{meng2025text2vdmtextvectordisplacement}, \brian{or deformation fields using neural representations \cite{eisenberger2021neuromorph, hanocka2018alignet, Maesumi23, Uy21}}.
There is also a large family of 3D editing methods built on implicit shape representations such as NeRF, Gaussian splatting, occupancy fields, and signed distance fields \brian{\cite{Barda24, Chen24a, mikaeili2023skedsketchguidedtextbased3d, chen2024generic3ddiffusionadapter, park2024ednerf, sabat2024nerfinsert3dlocalediting, nerfediting, controlnerf, Park2021nerfies}}.

\smallskip
\noindent \textbf{Neural Jacobian Fields.}
Neural Jacobian Fields (NJF) \cite{aigerman2022neural} pioneered in connecting classical differential deformation methods such as \citet{yu2004poisson} to neural, differentiable pipelines with geometric or semantic losses. This approach has powered compelling results for applications including UV mapping, re-posing, handle-based deformation~\cite{yoo2024plausible}, and unsupervised shape morphing and correspondence~\cite{sundararaman2024deformation}.

In the NJF approach, given a local transformation matrix $\Mm_k$ per face $k$, the least-squares best fit deformed vertices $\Phi^*$ to these differential transforms can be found by solving a Poisson equation with the cotangent Laplacian:
{\small\begin{equation}
\label{eq:njfpoisson}
    \Phi^* = \argmin_\Phi \sum_{k \in \text{all faces}} {a_k} \| \Phi \nabla_k^\top - \Mm_k\|^2_2 = L^{-1}\AA \nabla^\top \Mm
\end{equation}
}
\noindent where $\Mm$ is all $\Mm_k$ stacked, $a_k$ is the area of face $k$, $\nabla$ is the gradient operator, and $\mathcal A$ is the face mass matrix.
This solve is differentiable with respect to $\Mm$.
We note that the global step to optimize ARAP energy \cite{sorkine2007arap} is also a Poisson equation with the cotangent Laplacian (\cref{eq:global-solve}). As such, in \ourtechnique{}, we can use a similar differentiable solver while taking advantage of the regularization inherent in ARAP.

Neural methods that use NJF's differential deformation for text-based deformation include TextDeformer \cite{gao2023textdeformer} and MeshUp \cite{kim2024meshup} which optimize jacobians to deform a source shape into a different semantic target \eg turning a dog into a frog via a text prompt. Their deformations are not sufficiently restricted by construction and require an extra L2 loss between identity and the estimated jacobians to prevent losing the shape identity altogether. In contrast, our deformation framework is more contained by construction (\cref{sec:local-rots}), preserving the source details and updating the geometry to the desired style (see \cref{fig:comparison} and \cref{sec:evaluation}).

%% file: sections/3_method.tex
\section{Method} \label{sec:method}

Our method takes as input a source triangle mesh $\MM = (\VV, \FF)$ and a text prompt $\xx$. Our goal is to obtain a deformed mesh $\MM^* = (\VV^*, \FF)$ that semantically matches the style indicated by $\xx$.

We find this deformation by optimizing per-vertex unit normals of $\MM$.
\namanhCAMRDY{For a mesh with $|\VV|$ vertices, gradient descent directly optimizes a $|\VV|\times 3$ array of real numbers, \ie, a 3-element vector per vertex. These per-vertex vectors $\hat \UU$ are treated as \emph{target normals} used to solve for per-vertex rotations $\hat \RR$, and then, the deformed positions $\hat \VV$.}
\namanhCAMRDY{Specifically, }from a current estimate $\hat \UU = \{\hat \uu_k \mid k \in \{1\dots|\VV|\}\}$ of target normals, we first perform a \emph{local step} that obtains per-vertex rotation matrices $\hat \RR_k = \{\hat \Rr_k \in \SO(3) \mid k \in \{1 \dots |\VV|\}\}$ from the normals $\hat \UU$ (\cref{sec:local-rots}), followed by a \emph{global step} that obtains the deformed vertex locations $\hat \VV$ from the per-vertex rotations $\hat \RR$ (\cref{sec:global-solve}).

We refer to this pair of local step and global step as \emph{\ourtechnique{}}.
\ourtechnique{} is closely inspired by the multiple alternating local-global iterations of classical ARAP optimizations (which are repeated many times until convergence) \cite{sorkine2007arap}, but here condensed into \emph{a single local step and single global step} as a differentiable module, usable in a neural optimization or deep learning pipeline.
While for classical deformation applications, ARAP is run until convergence to achieve satisfactory results, in the context of our pipeline (\cref{fig:overview}), \ourtechnique{} running only one iteration is sufficient and offers the benefits of efficient differentiability.

\subsection{Local Rotations from Normals}
\label{sec:local-rots}

\begin{figure}[t]
    \centering
    \includegraphics[width=0.99\linewidth]{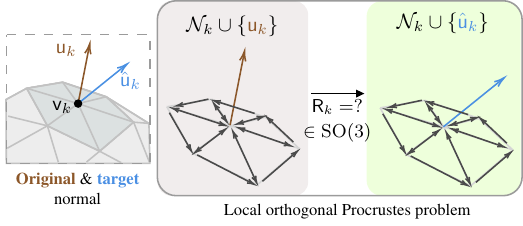}
    \vspace{-5pt}
    \caption{\textbf{Local Orthogonal Procrustes}. The single-iteration local step of our \ourtechnique{} energy solves for the best fit rotation given the original and and target normal.
    }
    
    \label{fig:procrustes}
\end{figure}

For a vertex $k$ with edge neighborhood $\NN_k$, with current estimated target vector $\hat \uu_k$ \namanhCAMRDY{(normalized to unit length)} and original normal vector $\uu_k$ (the area-weighted unit normal of vertex $k$ of the undeformed mesh), we compute a best fit rotation that transforms the bundle of vectors $\NN_k \cup \{\uu_k\}$ to the bundle $\NN_k \cup \{\hat \uu_k\}$ (see \cref{fig:procrustes}).
The best fit rotation $\hat \Rr_k$ minimizes the ARAP energy assuming fixed vertices $\hat \vv_k$, \ie 
{\begin{equation} \begin{split}\label{eq:arap-energy-local}
\hat \Rr_k = \argmin_{\Rr_k}
&\sum_{(i,j) \in \NN_k} w_{ij} \| \Rr_k \ee_{ij} - \ee_{ij} \|_2^2
\\
&\quad + \lambda a_k \| \Rr_k \uu_k - \hat \uu_k \|_2^2
\end{split}\end{equation}}
\noindent where $a_k$ is the Voronoi mass of vertex $k$; $\lambda$ is a hyperparameter that scales the strength of the rotation matching the source to the target normal; $\ee_{ij} \in \NN_k$ are all the edge vectors in the neighborhood of vertex $k$, and $w_{ij}$ are the cotangent weights~\cite{pinkall1993cotweights} of these edges.
Like \citet{liu2021normal}, we choose the spokes-and-rims neighborhood, consisting of halfedges in the vertex 1-ring, their twin halfedges, and halfedges opposite the vertex \cite{chao2010spokesandrims}.

This minimization is the \emph{orthogonal Procrustes problem} for each neighborhood, and can be solved \cite{liu2021normal} by finding 
\def\Ee{\mathsf E}
\def\Xx{\mathsf X}
\def\Ww{\mathsf W}
\begin{equation}
\Xx_k = \begin{bmatrix}
\Ee_k & \uu_k
\end{bmatrix}
\begin{bmatrix}
\Ww_k & \\ & \lambda a_k
\end{bmatrix}
\begin{bmatrix}
\Ee_k^\top \\ \hat \uu_k^\top
\end{bmatrix}
\end{equation}
where $\Ee_k$ is a $3 \times |\NN_k|$ matrix whose columns are the undeformed edge vectors in $\NN_k$, and $\Ww_k$ is a $|\NN_k| \times |\NN_k|$ diagonal matrix with the cotangent weights of the $\NN_k$ edges as the entries. (The $|\NN_k|$ dimensions in these matrices can be zero-padded to $\max_{k \in \{1\dots |\VV|\}} |\NN_k| $ for batched solutions.) Taking the SVD of $\Xx_k$, we can find $\hat\Rr_k$ (up to multiplying the last column of $\mathbf U_k$ by $-1$ to ensure $\det(\Rr_k)>0$) as
\begin{equation}
\begin{split}
\mathbf U_k \mathbf \Sigma_k \mathbf V_k^\top =\Xx_k \\
\hat\Rr_k = \mathbf V_k \mathbf U_k^\top
\end{split}
\end{equation}
\begin{figure}[!t]
    \centering
    \includegraphics[width=0.83\linewidth]{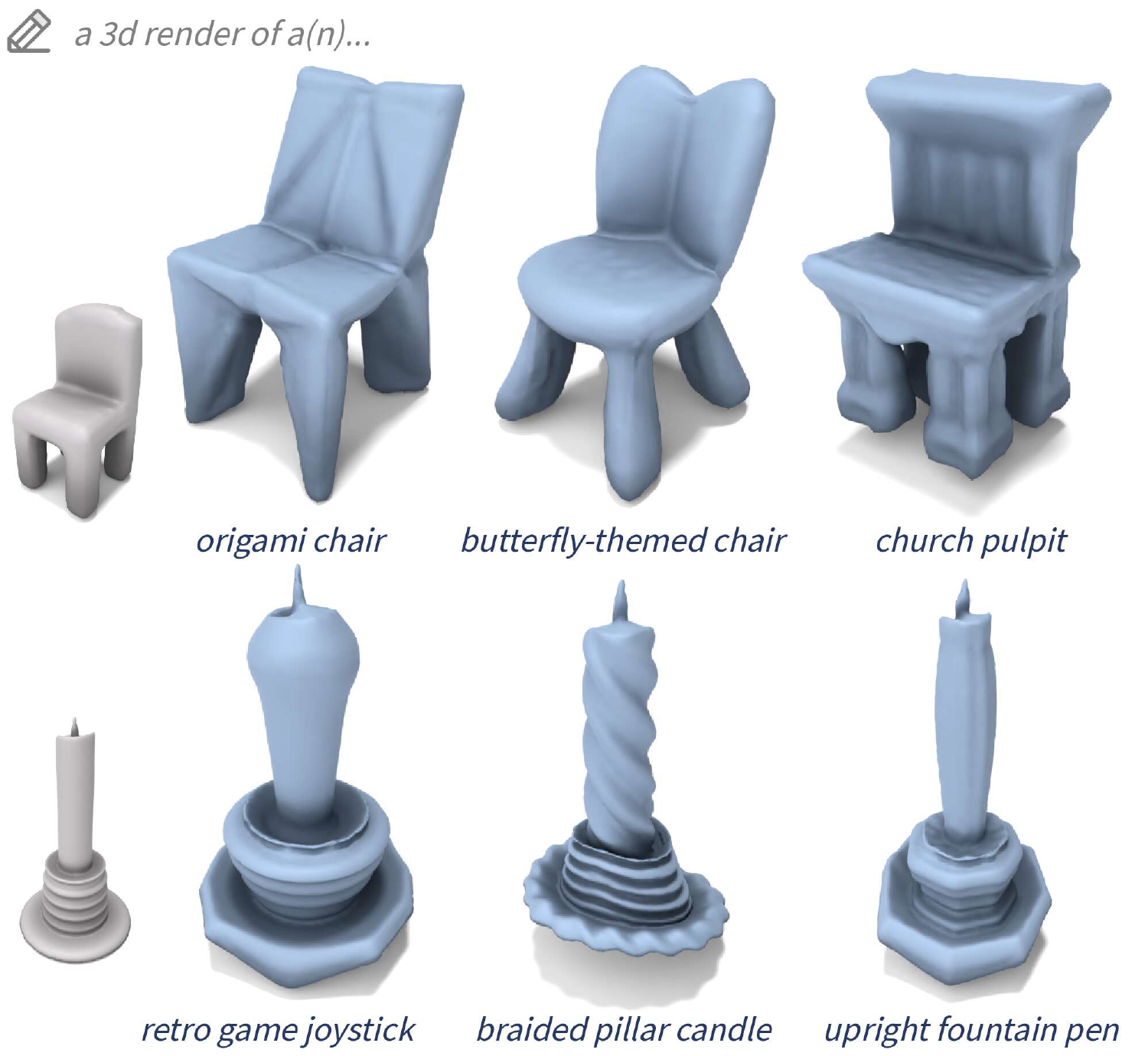}
    \vspace{-5pt}
    \caption{Our method is capable of deforming the same mesh towards different text-specified styles.}
    \label{fig:same-shape}
\end{figure}

Note that in classical ARAP iterative optimization, the term in \eqref{eq:arap-energy-local} is normally $\sum_{\ee_{ij} \in \NN_k} \left(w_{ij} \| \Rr_k \ee_{ij} - \ee'_{ij} \|_2^2\right)$ where $\ee'_{ij}$ is the vector of the most recent deformed edge $(i,j)$ in $\NN_k$ output by the previous ARAP optimization iteration.
Since \ourtechnique{} is meant as a differentiable module in a larger optimization process or learning pipeline, we condense the typically many local-global alternating steps of classical ARAP to just \emph{one local step and one global step}, hence the identification of $\ee'_{ij}$ with $\ee_{ij}$.
Coupled with setting $\lambda$ to an appropriately large value, which scales the strength of the rotation towards the requested normal $\hat \uu_k$, our \emph{single} local step is still able to achieve the required expressiveness and strength to make detailed deformations, yet regularized by the Procrustes solve to retain shape identity \emph{without} requiring an extra identity regularization loss as in \citet{kim2024meshup} and \citet{gao2023textdeformer}.

Note also that in NJF-based methods such as \cite{gao2023textdeformer, kim2024meshup}, the local step would be the identity function; a jacobian matrix per face is assumed given or predicted from upstream components. In our case, a matrix (a rotation) is not given, but computed from the \emph{target normal vector} for each element.

\subsection{Global Solve from Local Rotations}
\label{sec:global-solve}

Having obtained a rotation per neighborhood with our local step, we minimize the energy fixing the rotation matrices and solving for deformed vertex locations, i.e., finding the deformed vertices $\hat \VV$ such that
{\begin{equation}
\hat\VV = \argmin_{\tilde\VV}\sum_{k \in \{1\dots |\VV|\}} \sum_{(i,j) \in \NN_k} w_{ij}\| \Rr_k \ee_{ij} - \tilde \ee_{ij} \|_2^2
\end{equation}}
where $\ee_{ij} = (\vv_j - \vv_i)$, $\tilde \ee_{ij} = (\tilde\vv_j - \tilde\vv_i)$, and $w_{ij}$ is the cotangent weight of edge $(i,j)$. This is a linear least squares optimization for $\hat \VV$. As such, for the spokes-and-rims neighborhood, taking the derivative with respect to $\hat \VV$ and setting it to zero yields a linear equation in $\hat \VV$
\begin{equation}\label{eq:global-solve}
L\hat\VV = 
\begin{bmatrix}
\operatorname{rhs}(1)^\top \\
\vdots\\
\operatorname{rhs}(|\VV|)^\top
\end{bmatrix}
\end{equation}
{\small\begin{equation}
\operatorname{rhs}(k)= \!\! \sum_{\scriptscriptstyle (k,m,n)\in \NN^F_k}{ \!\! \frac{\Rr_k + \Rr_m + \Rr_n}{3} \left(\frac{w_{km}}2\ee_{km} + \frac{w_{kn}}2 \ee_{kn} \right) }
\end{equation}}
where $\NN_k^F$ are the faces adjacent to vertex $k$, each one having vertices $(k,m,n)$ (\ie even permutation such that $k$ is in front); $w_{km}, w_{kn} $ are the (undirected) cotangent weights of edges $(k,m), (k,n)$; and $L$ is the cotangent Laplacian.

Equation \eqref{eq:global-solve} is a Poisson equation.
As such, we can use the same solving and pre-factorization techniques as in NJF \cite{aigerman2022neural} to make our global solve step differentiable and efficient.

\subsection{Optimization Using a Semantic Visual Loss}
\label{sec:render-vis-loss}

\paragraph{Differentiable renderer and semantic loss.}
Our optimization is guided by a powerful pretrained text-to-image (T2I) diffusion model. We render the deformed mesh from multiple views using a differentiable rasterizer~\cite{laine2020nvdiffrast}, then feed the rendered views into a semantic visual loss, in our case Cascaded Score Distillation (CSD) \cite{decatur2024paintbrushcsd} using stages 1 and 2 of the image diffusion model DeepFloyd IF \cite{stabilityai2023deepfloydif}.

%% file: sections/4_experiments.tex
\section{Experiments} \label{sec:experiments}

\smallskip
\noindent \textbf{Optimization details. }
\label{sec:experiments-details}
Our initial guess for the target normals $\hat\UU$ is $\UU$, the original area-weighted vertex normals of the undeformed mesh $\MM$. We set the local step hyperparameter $\lambda=8$ (\cref{sec:local-rots}) for all of our optimizations. We run our optimization for 2500 epochs at a constant learning rate of $0.002$ using the Adam optimizer, each epoch being a batch of 8 renders fed to CSD loss. A full optimization run takes about 2 hours 15 minutes using a single A40 GPU. We remesh our source meshes for better behavior with the cotangent Laplacian (see \cref{sec:limitations}.) Further details on view sampling settings, CSD configuration, and source mesh preprocessing can be found in the supplementary material.

\subsection{Properties of \ourmethod{}} \label{sec:properties}

\paragraph{Generality and expressivity.} Our method is highly versatile, and is able to deform meshes from varied domains towards a wide range of styles  (\cref{fig:teaser,fig:gallery}).
Our method handles organic and articulated surfaces, such as animals and the human body, as well as man-made objects with sharp features and complex topology such as chairs.
The target style is specified by an open-vocabulary text prompt and can thus be described flexibly and intuitively.

The stylization manifests in a part-aware manner, conforming to the shape's geometry.
For the \emph{pineapple-themed} vase in \cref{fig:teaser}, our method adds a pineapple-like \emph{geometric texture} to the vase's body, while the vase's head is deformed with a different ripple pattern to resemble a pineapple head. For a human body, our method creates \emph{geometric details} to reflect a knight's armor in the appropriate locations, such as shoulder pads, a large chest plate, a crease across the waistline, and a hat on the head.
In \cref{fig:gallery}, the \emph{antique}, \emph{gothic}, and \emph{cardboard} chairs' styles are reflected by both local geometric details and the silhouette of the deformed mesh.

\begin{wrapfigure}{r}{0.342\linewidth}
\centering
\includegraphics[width=1.0\linewidth, trim=3.0cm 4cm 2.5cm 2.3cm]{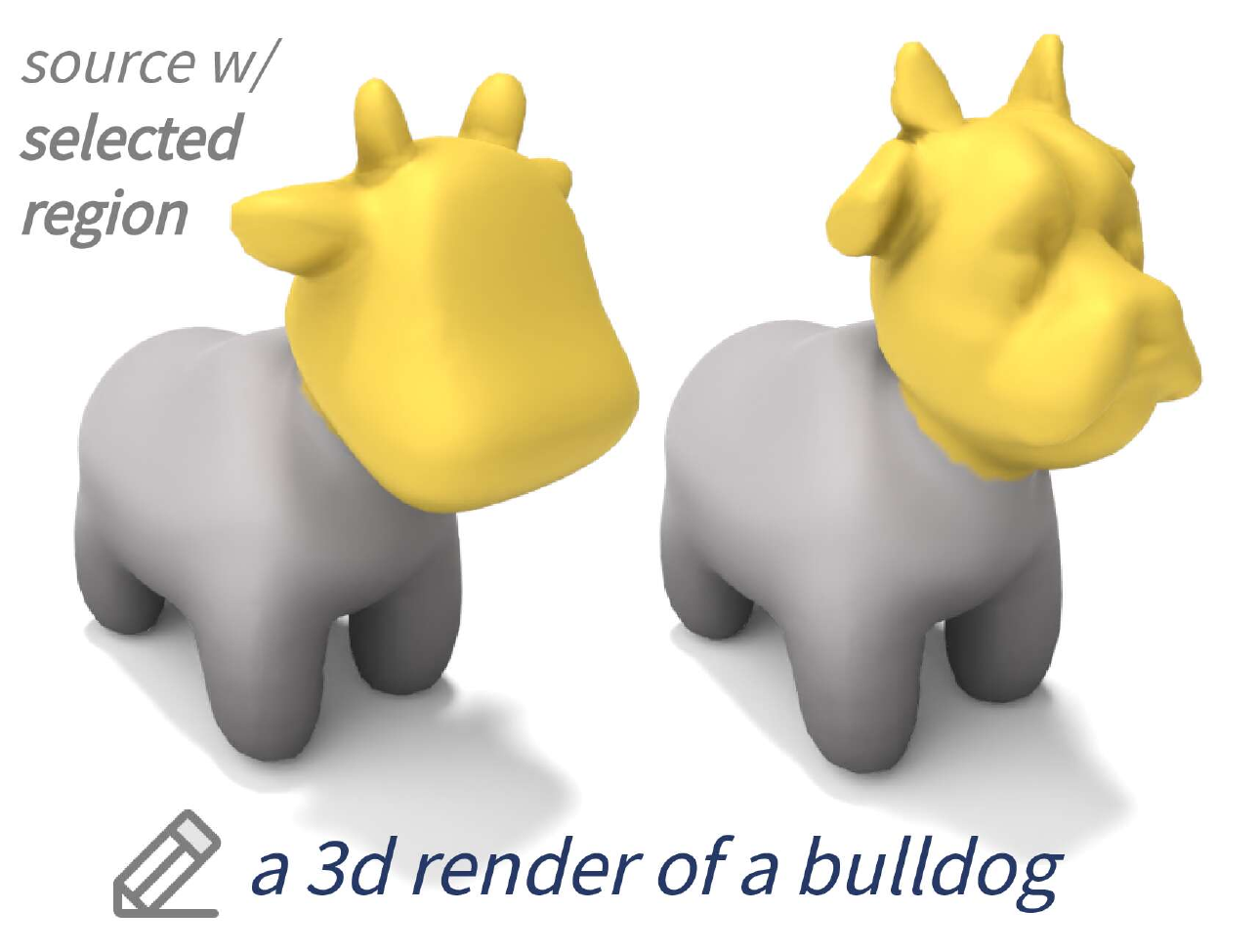}
\label{fig:local-deform}
\end{wrapfigure}
We can further take advantage of part awareness to stylize \emph{only} select regions. In the inset, we localize the deformation to the head by setting the rotation of vertices outside the region to the identity matrix every iteration. The deformation is contained within the local region, yet detailed and appropriate for the rest of the body. We observe no boundary artifacts, showing dARAP's beneficial regularizing effects.

\begin{figure}[!t]
    \centering
    \includegraphics[width=0.93\linewidth]{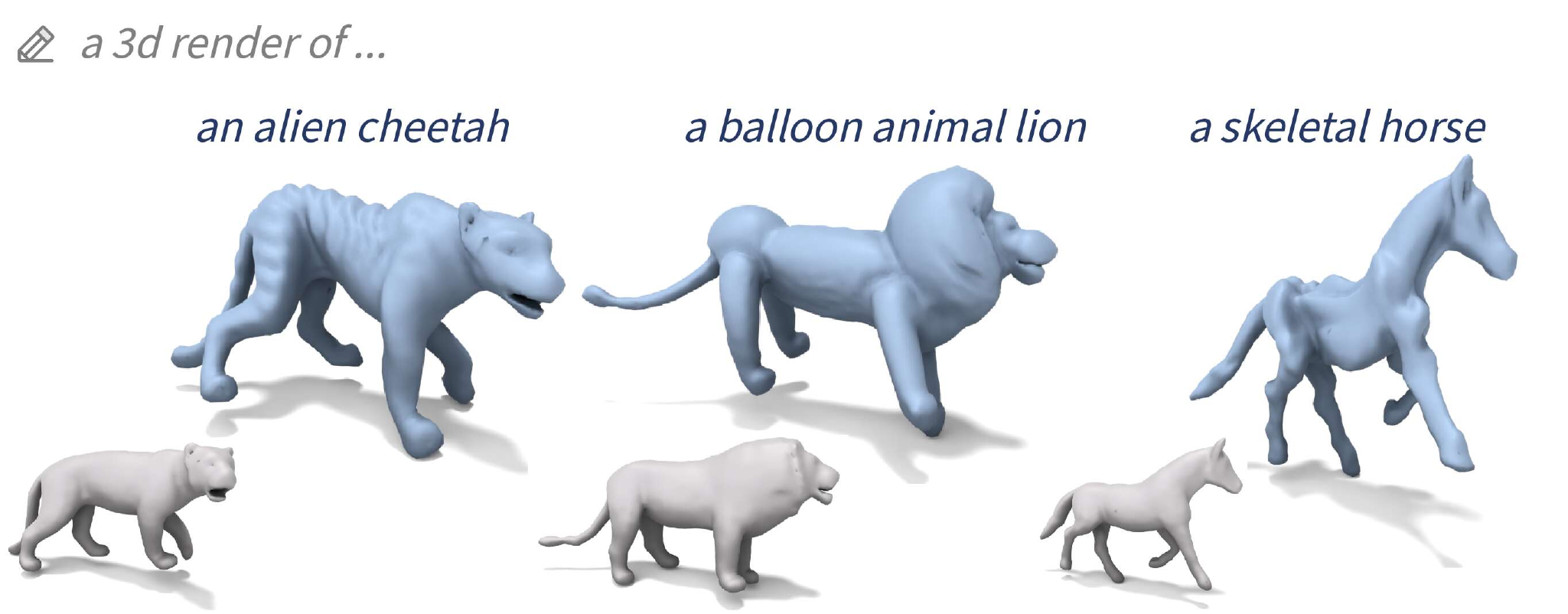}
    \vspace{-5pt}
    \caption{\textbf{Pose preservation.} \ourmethod{} stylize the shape according to the text prompt, while decently keeping characteristics of the source shape like the relative positioning of the limbs and the angle of the head.}
    \label{fig:pose-preservation}
\end{figure}

\smallskip
\noindent \textbf{Identity preservation.} Our method applies the prescribed style to the input mesh expressively while preserving important characteristics. In \cref{fig:pose-preservation}, each input animal has a unique pose, \eg the folded front leg of the horse. The deformation recognizably keeps the pose while stylizing the body towards the \emph{skeletal} style. Similarly, the stylization of the person shape from \cref{fig:teaser} maintains body proportions. Additionally, our method preserves other distinct shape properties, \eg, the animal's facial features (\cref{fig:pose-preservation}) and the chairs' parts (\cref{fig:gallery}), with semantic correspondences maintained (\cref{fig:correspondence_fig}).

\begin{figure}
    \centering
    \includegraphics[width=0.99\linewidth]{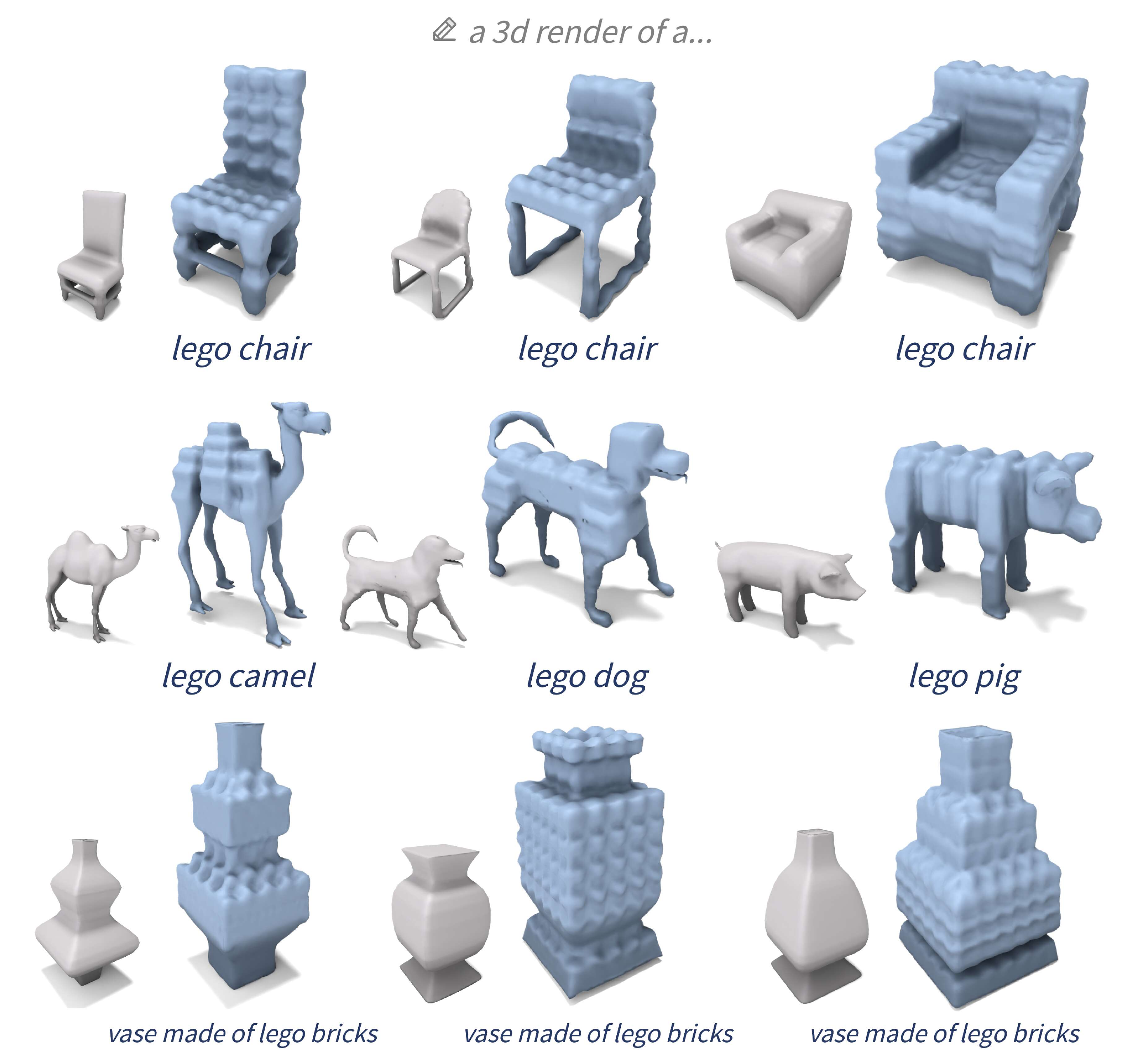}
    \vspace{-5pt}
    \caption{Our method is capable of deforming \textit{different} input meshes towards the same text-specified style. Even with the same prompt in a shape class (\emph{a 3d render of a lego chair}, \emph{a 3d render of a vase made of lego bricks}), different starting shapes (different chairs, different vases) result in stylizations that closely align to the identity of the original shape, while still strong enough to induce more blocky components and \emph{lego}-like surface textures.}
    \label{fig:samestyle-lego}
\end{figure}

We attribute our pronounced yet identity-preserving stylizations to our normal-based deformation representation. 
A pretrained vision foundation model provides strong guidance toward the style prompt but can also easily impart deformations that significantly alter the identity of the shape, as witnessed in previous work that used jacobian-based deformations \cite{gao2023textdeformer, kim2024meshup}, see \cref{fig:comparison,fig:pineapple-lamp}. By contrast, our deformation is driven by \emph{rotations} of normals, further regularized as a best-fit rotation over the spokes-and-rims neighborhood (\cref{sec:local-rots}.) This formulation selects for local rigid changes that are discouraged from scaling or shearing, thus preventing excessive structural changes while requiring no extra identity regularization loss on the local transforms.

\smallskip
\noindent \textbf{Specificity.} Our method performs diverse shape stylizations that adhere to the target style prompt with high detail. In \cref{fig:same-shape}, we deform the source shape into different styles: the \emph{origami} chair's backrest is thin and has creases as with paper folds; the \emph{church pulpit} style is thicker with overhangs appropriate of church furniture. As further seen by the detailed prompts and styles in \cref{fig:teaser,fig:gallery}, our method produces distinct styles and shows granular effectiveness.

\smallskip
\noindent \textbf{Robustness.} Our method exhibits robustness across shape categories and instances within the category. In \cref{fig:samestyle-lego}, the same \emph{lego} style is applied to chairs, animals, and vases. Each domain has unique geometry: the chairs have varying parts (\eg, the types of legs and backrests), the animals have smooth geometry, and the vases have sharp edges and rotational symmetry. Still, our method consistently conveys the style on the source shape with a \emph{lego} brick-like surface pattern and by cubifying the geometry.

\begin{figure}[!t]
    \centering
    \includegraphics[width=0.99\linewidth]{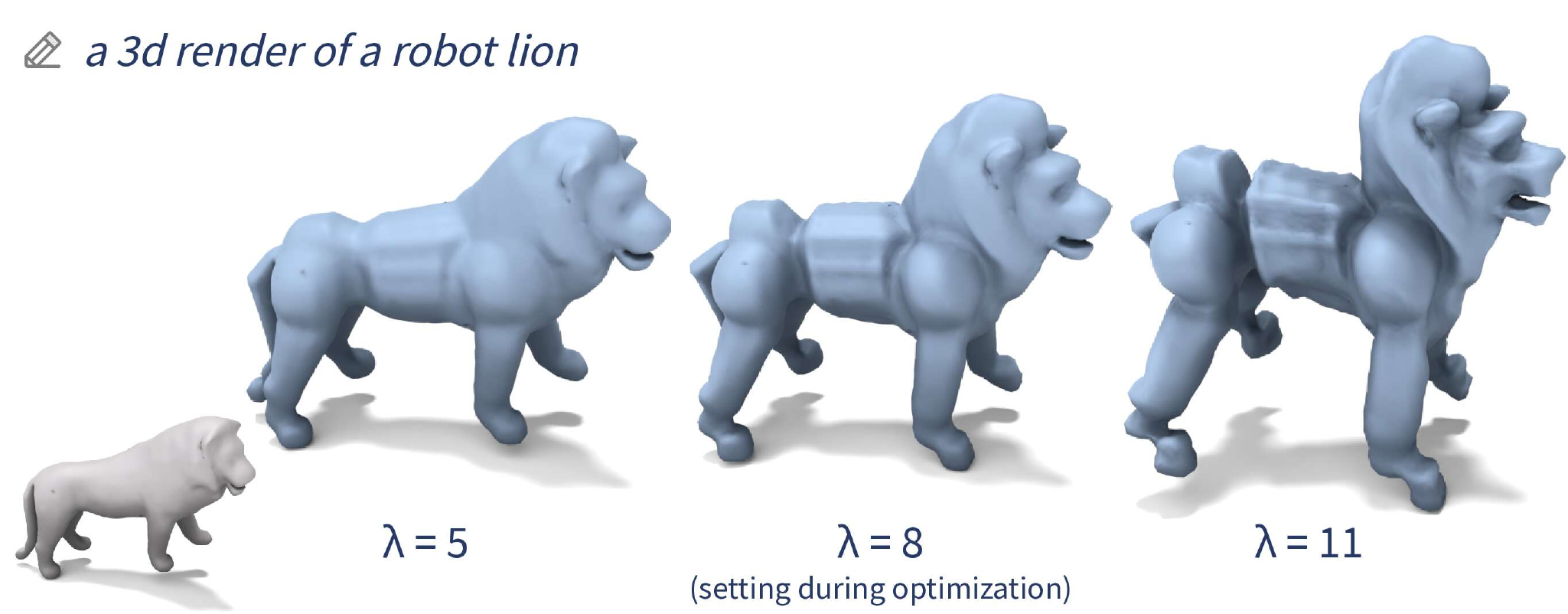}
    \vspace{-5pt}
    \caption{\textbf{Changing stylization strength after optimization.} Normals found by optimization using $\lambda = 8$ can be conveniently re-applied after optimization using a different $\lambda$ to tune the stylization strength on demand. Both larger and smaller $\lambda$ result in salient and clean stylizations at the required strength.}
    \label{fig:lambdatweak}
\end{figure}

\smallskip
\noindent \textbf{Tunable stylization strength.} A specific value of $\lambda$ is used by deformations during the optimization process. However, this $\lambda$ can be changed when re-applying the saved optimized normals to the source mesh, allowing \textit{user control} separate from the optimization pipeline itself, as shown in \cref{fig:lambdatweak}. Recall from \cref{eq:arap-energy-local} that the hyperparameter $\lambda$ influences the match from the original vertex normal to the target normal. 
As seen in \cref{fig:lambdatweak}, as the value of $\lambda$ increases, the geometry of the deformed shape is sharpened to strengthen to ``robot" style effect, while a lower $\lambda$ results in deformed normals being closer to the original ones, though the desired style is still visible.
Notably, we observe that using an inference $\lambda$ value larger than that used during optimization results in a more geometrically salient yet still sensible stylization, further demonstrating the robustness of our method.

\subsection{Evaluation} \label{sec:evaluation}
We contrast our method to two recent text-guided mesh deformation methods, TextDeformer \cite{gao2023textdeformer} and MeshUp \cite{kim2024meshup}
using public code released by the authors.

\begin{figure}[!b]
    \centering
    \includegraphics[width=0.96\linewidth, trim=0 0 0 80]{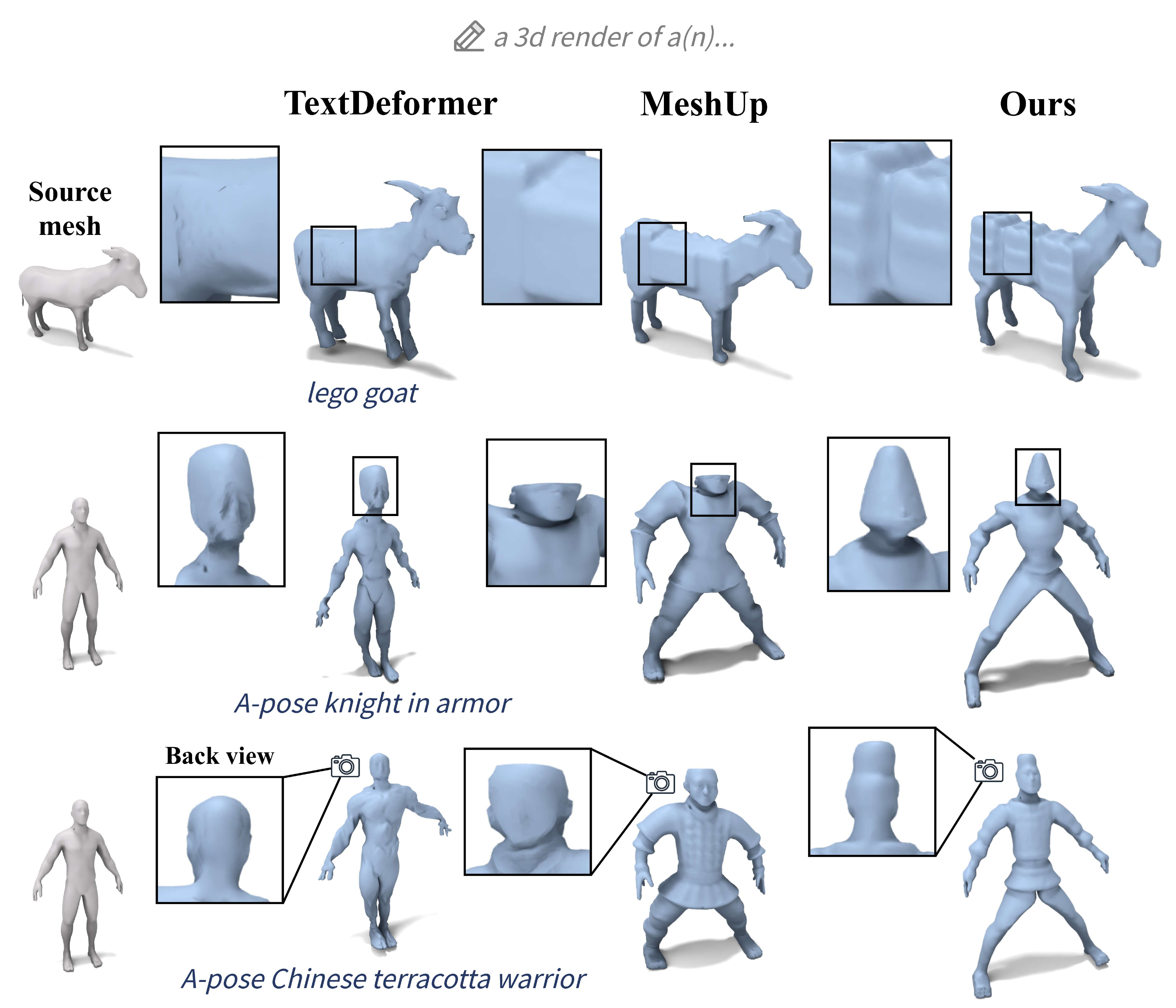}
    \vspace{-5pt}
    \caption{\textbf{Comparison with baselines.} We compare our method with the alternative deformation techniques TextDeformer \cite{gao2023textdeformer} and MeshUp \cite{kim2024meshup}.
    While the baseline methods have a weaker stylization effect, change the poses, or create geometric artifacts on some examples, \ourmethod{} cleanly achieves both the desired style and remains faithful to the shape of the source meshes.}
    \label{fig:comparison}
\end{figure}

\smallskip
\noindent \textbf{Qualitative comparison.} In \cref{fig:comparison}, we show deformation results for TextDeformer~\cite{gao2023textdeformer}, MeshUp~\cite{kim2024meshup}, and our method. For comparison, all three methods use the same text prompt, source mesh, and view sampling settings. 
TextDeformer distorts the surface, changes the pose of the source shape, and does not achieve the target style. MeshUp does stylize following the text prompt, but in some cases, its surface texture may be weaker than ours, as seen in the \emph{lego} goat example.

In other cases, MeshUp's stylization is strong but produces notable distortion in arms and body proportions: in the examples \emph{knight in armor} and \emph{Chinese terracotta warrior}, the body gets a broader, stouter stature and does not preserve the identity and body proportions of the source shape as well as our method does. Moreover, our method does not encounter certain artifacts that MeshUp introduces, like the crushed head of the \emph{knight in armor} example, or the Janus effect with a duplicate face and chestplate on the back for \emph{Chinese terracotta warrior.} We achieve a prominent desired style with fewer artifacts and better preservation of source shape features. \namanhCAMRDY{More qualitative examples are given in the supplementary material.}

\smallskip
\noindent \paragraph{Quantitative results.} As a proxy for evaluating identity preservation, we measure the mean and standard deviation of the ratio (deformed triangle area / original triangle area) summarized over all triangles across the chosen meshes. We use 20 mesh-prompt pairs chosen from results seen throughout the paper; per-mesh triangle counts range from 10 thousand to 20 thousand. Source meshes are normalized to fit a side-length-2 cube centered at the origin; deformed meshes are normalized to have the same bounding box diagonal length as the source, as is done after the Poisson solve of MeshUp and our method.
This quantity measures distortion and estimates how well the bounding box is respected: deformations that shrink the mesh lead to a larger rescale factor during the bounding-box-restoring normalization, thus inflating the face area ratio, and vice versa.
Ideal identity-preserving values are ratio 1 with 0 standard deviation.

\Cref{tab:tri-area-ratio} summarizes the quantitative comparison with the baseline methods \cite{gao2023textdeformer, kim2024meshup}. These deformations use jacobians, which are more prone to changing triangle scale and compromising the integrity of the mesh. In contrast, we represent a target deformation by surface normals, coupled with our \ourtechnique{} layer that regularizes the resulting deformation and better preserves the original triangle area, improving faithfulness to the input shape.
Indeed, as \cref{tab:tri-area-ratio} shows, our triangle area ratio has an average closer to 1 with a lower standard deviation than MeshUp and TextDeformer. \namanhCAMRDY{We also include a quantitative evaluation of CLIP similarity to the prompt for these three methods on the same shapes (see the supplementary material); our method achieves better CLIP similarity to the prompt.}

\begin{table}[t!]
\small
\centering
\begin{tabular}{lcc}
\toprule
Method  & Ratio mean & Ratio std. dev. \\
\midrule
TextDeformer \cite{gao2023textdeformer} & 0.827 & 0.360  \\
MeshUp \cite{kim2024meshup} & 1.288 & 0.363 \\
\ourmethod{} (\textbf{ours}) & \textbf{1.080} & \textbf{0.233} \\
\bottomrule
\end{tabular}
\caption{\textbf{Triangle area preservation.} As a surrogate for measuring identity preservation, we compute the mean and standard deviation of the triangle area ratio between the deformed and source shapes. Our method preserves the triangle area better than the other methods, with an \textbf{average ratio closer to 1} and a lower standard deviation.}
\label{tab:tri-area-ratio}
\end{table}

\subsection{Limitations} \label{sec:limitations}

Our method uses the cotangent Laplacian, which works well only on manifold meshes and is sensitive to triangle aspect ratios.
This is also true for other methods that use the cotangent Laplacian, such as TextDeformer~\cite{gao2023textdeformer} and MeshUp~\cite{kim2024meshup}.
To mitigate this, we remesh input meshes with isotropic explicit remeshing~\cite{hoppe1993remesh}, optionally after manifold preprocessing  \cite{huang2018manifold}.

Another limitation is the possibility of self-intersection in deformed meshes (\cref{fig:pineapple-lamp}).
The rods of the source lamp are rotated towards the center and intersect each other.
This can be somewhat mitigated by adjusting the deformation strength parameter $\lambda$ after optimization, as discussed in \cref{fig:lambdatweak}, a strategy not straightforwardly available to MeshUp.

%% file: sections/5_conclusion.tex
\section{Conclusion} \label{sec:conclusion}

\begin{figure}[!t]
    \centering
    \vspace{-3pt}
    \includegraphics[width=0.94\linewidth,trim=0 0 0 5,clip]{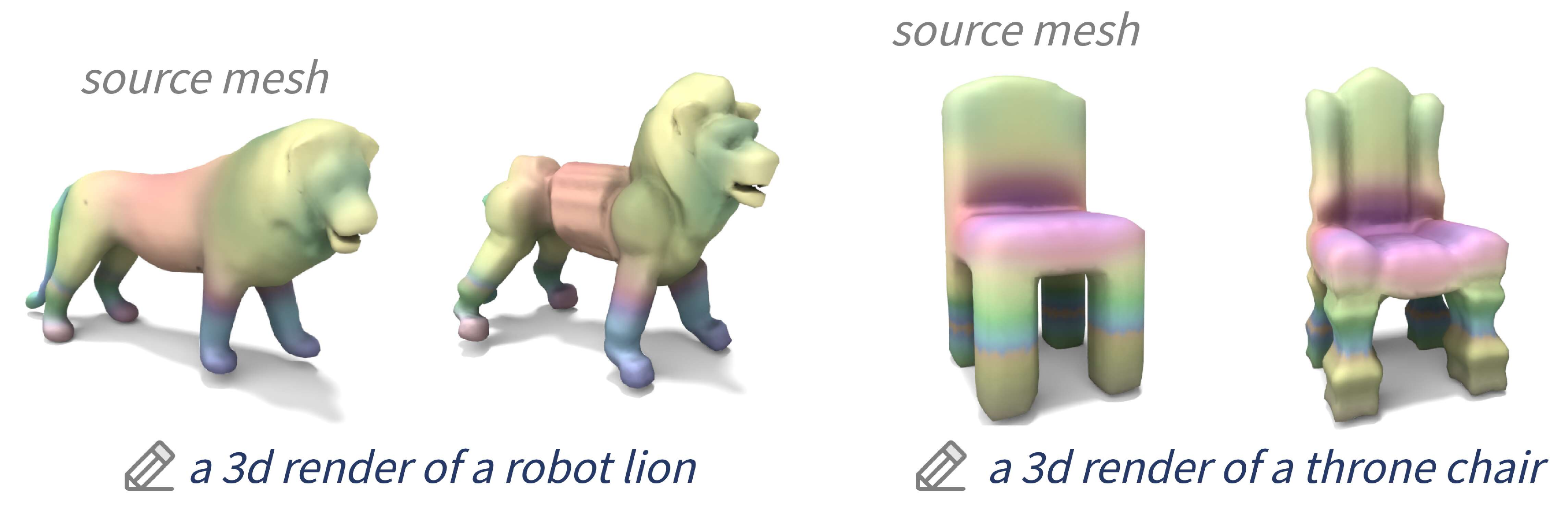}
    \vspace{-5pt}
    \caption{%
    \textbf{Correspondence.}
    The meshes deformed with our method preserve semantic correspondence to the source mesh.
    Deformed vertices have the same color as the corresponding source vertices, colored by the source shape's wave kernel signature.}
    \label{fig:correspondence_fig}
\end{figure}

\begin{figure}[!b]
    \centering
    \includegraphics[width=0.89\linewidth]{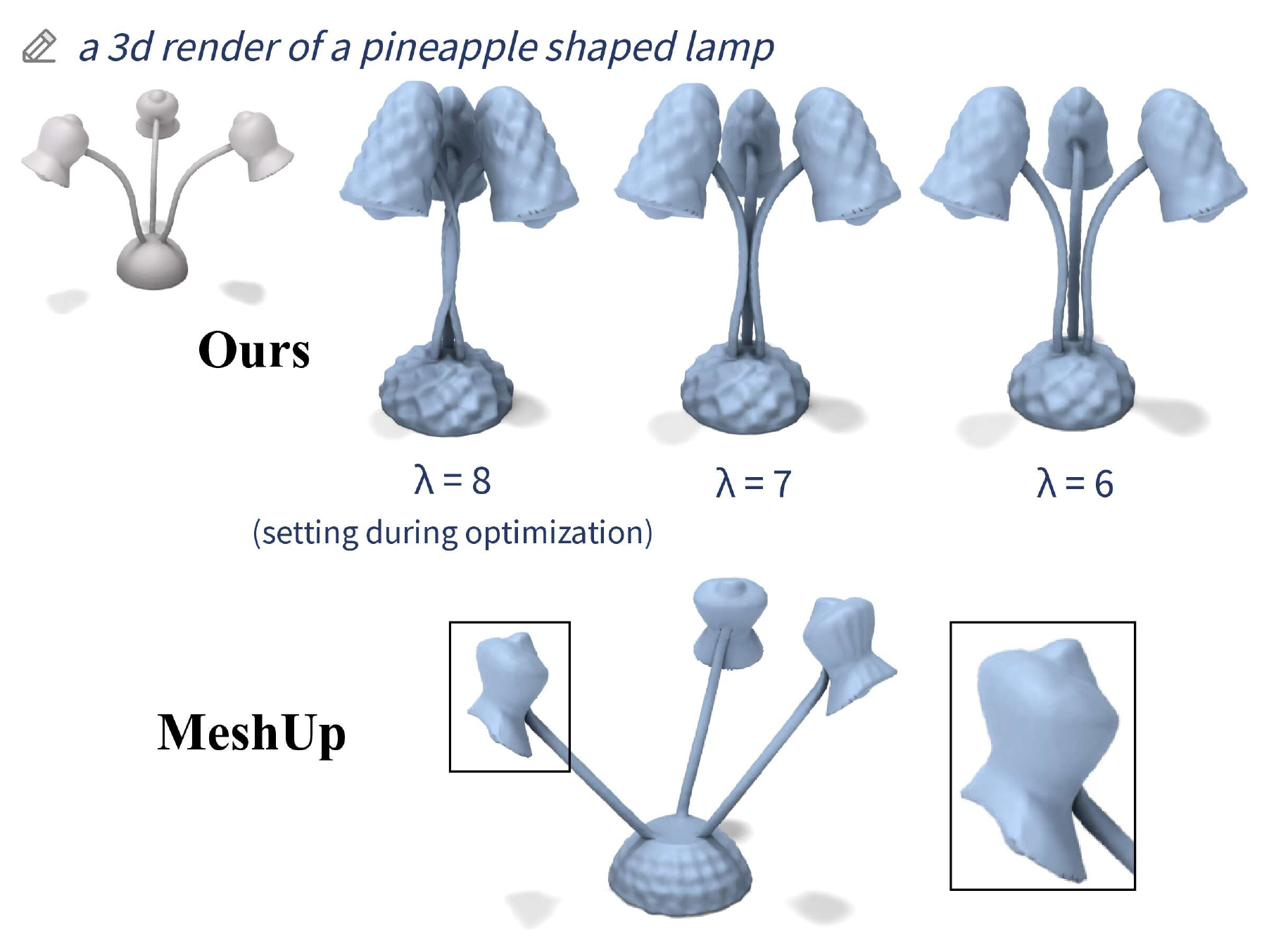}
    \vspace{-5pt}
    \caption{%
    \textbf{Limitations.}
    Our method may produce self-intersections.
    Decreasing the parameter $\lambda$ \textit{after} the optimization process can alleviate the self-intersection with only a mild reduction in the stylistic surface details.
    MeshUp warps the 3-lamp structure and the individual lamps and exhibits less geometric pineapple texture.}
    \label{fig:pineapple-lamp}
\end{figure}

In this work, we presented \ourmethod{},
a technique for deforming meshes to achieve a text-specified style.
A key claim of the work is that prescribing a deformation via surface normals allows the recovery of deformed vertices that adhere well to the input geometry while still being expressive.
We demonstrate high-quality, detailed geometric stylizations that respect the input shape's identity.

As part of \ourmethod{}, we introduce \ourtechnique{}, a differentiable neural network layer that deforms a surface to achieve target normals.
Our layer is simple yet effective: while traditional applications require iterating ARAP to convergence for a desirable solution, we find that within our neural network pipeline, good results can be achieved with \emph{only a single step}.
We speculate that by iteratively updating target normals through gradient descent, we can avoid (in a \ourtechnique{} forward pass) the standard practice of needing to repeatedly iterate between local and global ARAP steps.  
Moreover, \ourtechnique{} is general and may be used for other geometry tasks where ARAP is useful, such as parameterization, re-posing, collisions, editing, and more.

In the future, we are interested in leveraging unsupervised segmentation strategies~\cite{decatur20233dhighlighter,decatur2024paintbrushcsd} to perform localized geometric stylization. In addition, while our method is topology-preserving, follow-up work could explore edits and deformations that add explicit parts or change topology.

\namanhCAMRDY{
\section{Acknowledgments}
We would like to thank Justin Solomon, Ana Dodik, Haochen Wang, Richard Liu, and members of the 3DL lab for their helpful comments, suggestions, and insightful discussions. We are grateful for the AI cluster resources, services, and staff expertise at the University of Chicago. This work was supported by BSF grant 2022363, NSF grants 2304481, 2241303, 2402894, and 2335493, and gifts from Snap Research, Adobe Research, and Google Research.
}